\documentclass[twocolumn,epsfig,superscriptaddress,prl]{revtex4-2}

\usepackage{graphicx}
\usepackage{amssymb}
\usepackage{amsmath}
\usepackage{hyperref}
\usepackage{changes}
\usepackage{comment}
\usepackage{mathtools, nccmath}
\usepackage[caption=false]{subfig}
\usepackage{ragged2e} % for the \justifying macro
\DeclareCaptionJustification{justified}{\justifying}
\usepackage{cancel}
\usepackage{stackengine}
\usepackage{placeins}
\usepackage[percent]{overpic}
\usepackage{multirow}
\usepackage{array}
\usepackage{booktabs}
\usepackage{pgfplots}
\pgfplotsset{compat=1.8}
\DeclareMathAlphabet{\mathpzc}{OT1}{pzc}{m}{it}
\usepackage{xcolor}

\begin{document}
\title{Origin of Subharmonic Gap Structure of DC Current-Biased Josephson Junctions}	

\author{Aritra Lahiri}
\email[]{aritra.lahiri@uni-wuerzburg.de}
\affiliation{Institute for Theoretical Physics and Astrophysics,
University of W\"urzburg, D-97074 W\"urzburg, Germany}
\author{Sang-Jun Choi}
\email[]{aletheia@kongju.ac.kr}
\affiliation{Department of Physics Education, Kongju National University, Gongju 32588, Republic of Korea}
\author{Björn Trauzettel}
\affiliation{Institute for Theoretical Physics and Astrophysics, University of W\"urzburg, D-97074 W\"urzburg, Germany}
\affiliation{W\"urzburg-Dresden Cluster of Excellence ct.qmat, Germany}
\date{\today}

\begin{abstract}
{The current-voltage characteristics of Josephson junctions exhibit a subharmonic gap structure (SGS), denoting jumps at specific voltages. While the prevalent multiple Andreev reflection theory matches the experimentally observed SGS, it is limited to a DC \emph{voltage} bias. For a DC \emph{current} bias, existing theories are restricted to low-transparency junctions and fail to capture the full SGS. We introduce a microscopic Floquet approach applicable for arbitrary transparencies, and recover the correct SGS for a DC \emph{current} bias. We provide a comprehensive understanding of SGS for a DC \emph{current} bias, which entails two-quasiparticle tunneling processes absent in existing theories, via two complementary perspectives: in the frequency domain, as generalised Andreev reflections absorbing multiple energies, and in the time domain, as the interference of non-equilibrium current pulses.}
\end{abstract}
\maketitle

Josephson junctions (JJs)~\cite{Josephson1962, Josephson1964, Josephson1965, Anderson1963, Rowell1963, Yanson1965}, central to various applications spanning metrology and quantum computing~\cite{Krantz2019}, exhibit a wide range of physical phenomena under different biasing conditions. However, theoretical studies focusing predominantly on phase- or voltage-biased JJs out of convenience  do not represent typical experiments where low impedance in the superconducting state~\cite{Likharev1979,Baronebook} often imposes a \emph{current bias}. Owing to their highly non-linear characteristics, the physical behaviour of JJs is strongly influened by the biasing conditions. Consequently, this discrepancy in biasing schemes hinders an accurate comparison of theory and experiment.

A key component of this discrepancy involves the subharmonic gap structure (SGS), jumps occuring in the current-voltage characteristics (IVC) at voltages equalling integer-fractions (subharmonics) of $2\Delta$, where $\Delta$ is the superconducting gap. There are primarily two models to address a DC current bias: First, the phenomenological Resistively and Capacitively Shunted Junction (RCSJ) model~\cite{McCumber1968,Stewart1968,Scott1970}, which sacrifices all microscopic details for simplicity, and naturally fails to capture any SGS. Second, a theory pioneered by Werthamer and Larkin (WL) ~\cite{Werthamer1966,Larkin1967} which incorporated microscopic details in DC current-biased JJs, but only in tunnel-type junctions with low transparencies. However, its prediction of SGS at mean voltages $\langle V\rangle=2\Delta/((2n-1)e)$ for integer $n$~\cite{Schulp1978b, Choi2022, McDonald1976, Schulp1978a, Zorin1983}, invoking the mechanism of Josephson self-coupling (JSC), is in discord with experiments which consistently observe subharmonics at $2\Delta/(ne)$ for all integers~\cite{Barnes1969, Hansen1973, Rowell1968, Kleinasser1994, Maezawa1994, Maezawa1995, Post1994, Scheer1997, Scheer1998, Ludoph2000, Naaman2001, Naaman2004, Naveh2001, Gul2017, Ridderbos2018, Barati2021}. While the Multiple Andreev Reflection (MAR) theory~\cite{Klapwijk1982, Blonder1982, Octavio1983, Bratus1995, Arnold1987, Averin1995, Cuevas1996, Cuevas2002, Cuevas2006} correctly predicts the experimentally observed SGS, it holds only for JJs which effectively experience a DC voltage bias, such as quantum point contacts or weak links with high junction resistance~\cite{Scheer1997, Scheer1998, wertnote}, in the presence of small shunt-resistances~\cite{Deacon2017,Liu2025}, or environmental impedance-matching~\cite{Chauvinthesis}. Notably, the MAR theory is not applicable for a DC current bias, leaving the discrepancy in biasing schemes unresolved.
%~\cite{Schrieffer1963, Hasselberg1974, vdPost1994}

We solve this long-standing problem, presenting a microscopic solution of DC current-biased JJs valid for arbitrary junction transparencies. For small tunnel coupling $\mathcal{T}$, we retrieve the WL theory~\cite{Werthamer1966,Larkin1967,McDonald1976} exhibiting odd subharmonics $e\langle V\rangle=2\Delta/(2n-1)$. With increasing $\mathcal{T}$, we provide a theoretical demonstration of SGS at \emph{all} $e\langle V\rangle=2\Delta/n$ for a DC current bias. We explain this phenomenon with two complementary approaches. While a DC voltage excites tunnelling quasiparticles by a single value equalling the voltage~\cite{Bratus1995}, a DC current bias generates an AC voltage, simultaneously exciting quasiparticles by multiple energies~\cite{McDonald1976}. In the leading correction to the WL current, two-quasiparticle tunnelling processes absorbing multiple energies enhance the resonance condition, and generate the even subharmonics. Moreover, in time domain, the AC voltage comprises a train of pulses, imbibing quasiparticles with a $\pi$ phase each~\cite{Choi2022}. The subharmonics arise from resonant addition of these currents, accounting for the phases. The phases cancel for two-quasiparticle processes beyond the WL limit, altering the resonance condition, and resulting in even subharmonics. Our formalism is not limited to $s$-wave superconductors, which we consider for simplicity.

\begin{figure*}[ht]
\includegraphics[width=\textwidth]{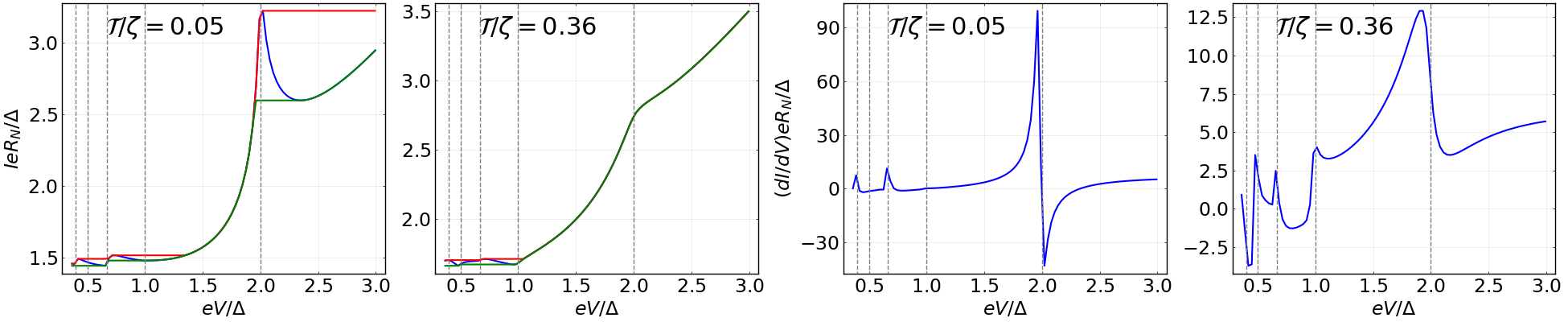}\rlap{\raisebox{1.55cm}{\hspace{-17.12cm}\includegraphics[width=0.1\textwidth]{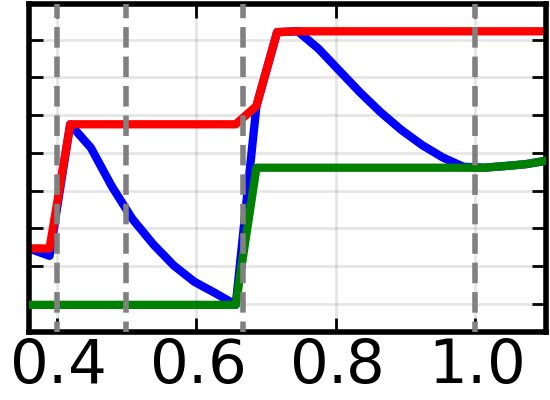}}}\rlap{\raisebox{1.55cm}{\hspace{-12.85cm}\includegraphics[width=0.1\textwidth]{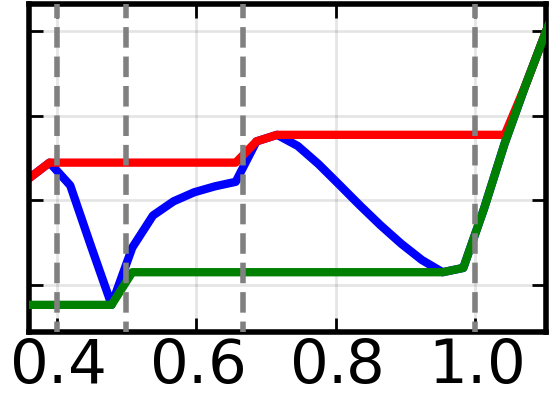}}}
\caption{Exact current-voltage characteristics (IVCs) using our non-perturbative numerical method in Eq. \eqref{If} and the corresponding differential conductance for a DC current bias with varying $\mathcal{T}$. In (a-b), the upper (red) and lower (green) envelopes of the current show the theoretically predicted hysteresis loops (inset: zoomed view of sub-gap region).  The even subharmonics $e\langle V\rangle=2\Delta/n$, which are absent for $\mathcal{T}/\zeta=0.05$, appear with increasing $\mathcal{T}$. Simultaneously, all subharmonic peaks are increasingly smoothened. Gray dashed lines mark the subharmonics $2\Delta/n$. The transparency is given by $4(\mathcal{T}/\zeta)^2/[1+(\mathcal{T}/\zeta)^2]^2$. We use $\Gamma=0.01\Delta$, and $\zeta=5\Delta$. The critical currents are shown in Appendix G. }
\label{Fig1} 
\end{figure*}

\textit{Model.}---We consider a single-channel Josephson junction, with two $s$-wave superconducting leads, connected to superconducting reservoirs~\cite{Cuevas1996}. A gauge transformation shifts the voltage into the tunnel couplings~\cite{Cuevas1996}. The Hamiltonian is then given by the sum of lead (L: left, R: right) and tunnel terms, $H=H_L+H_R+H_\mathcal{T}$, with,
\begin{equation}
\begin{split} H_{L/R}&={\textstyle\sum_{j\in L/R,\sigma}}-\zeta c_{j+1\sigma}^\dagger c_{j\sigma} + \Delta c^\dagger_{j\downarrow}c^\dagger_{j\uparrow} + \text{h.c.},\\
H_\mathcal{T}&={\textstyle\sum_\sigma} -\mathcal{T} W(t)c_{L(x=0)\sigma}^\dagger c_{R(x=0)\sigma} + \text{h.c.},
\end{split}\label{Ham}
\end{equation}
where $\zeta$ is the hopping amplitude, $\mathcal{T}$ is the tunnel coupling, and $W(t)=e^{-i\phi(t)/2}$ with $\phi(t)=2e\int_{-\infty}^t d\tau V(\tau)$ ($ \hbar=1$) being the Josephson phase. We consider the non-equilibrium steady state wherein the voltage is periodic $V(t+T_0)=V(t)$. From the Josephson relation $d\phi(t)/dt=2eV(t)$, it follows that $\phi(t+T_0)=\phi(t)+2\pi$~\cite{phinote} with period $T_0=\pi/(e\langle V\rangle)$, where $\langle V\rangle=\int_t^{t+T_0}V(\tau)d\tau/T_0$. Consequently, the Hamiltonian is periodic with period $T=2T_0=2\pi/(e\langle V\rangle)$.

This periodicity makes the problem amenable to the Floquet formalism~\cite{Martinez2003,Stefanucci2008,Gavensky2021,SanJose2013,Bolech2005,Gavenskythesis}. The tunnelling Hamiltonian has a non-trivial Floquet transform $H_{LR,m}=-\mathcal{T}\int_0^T W(t)e^{im\Omega t} dt/T=-\mathcal{T}[W_m(\tau_0+\tau_3)/2 + W_{-m}^*(\tau_0-\tau_3)/2]$ where $\Omega = e\langle V\rangle$ is the fundamental Floquet frequency, $W_m = \int_0^T e^{-i\phi(\tau)/2}e^{im\Omega \tau} d\tau /T$, and $\tau$ denotes the Pauli matrices in Nambu space. Since $W(t+T_0)=-W(t)$, only \emph{odd} harmonics $W_{(2n-1)}$ $\forall n\in\mathbb{Z}$ are non-zero~\cite{McDonald1976}. This implies that quasiparticles exchange energy in \emph{odd} multiples of $e\langle V\rangle$ upon tunnelling, with the amplitude $W_m$.  

The Floquet components $I_{k}$ of the current $I(t)=\sum_kI_ke^{-ik\Omega t}$ are obtained as~\cite{Lahiri2023,Cuevas1996,Gavensky2021,Gavenskythesis}
\begin{equation}
\begin{split}
I_{k}=&\sum_{m,n}e\begingroup\textstyle\int\endgroup\limits_0^\Omega d\omega \ \mathbf{Tr}\bigg[ \begin{matrix}\tau_3\Sigma_{T,LR,(n+k)m} G_{RL,mn}^{<}(\omega) \\ -(L\leftrightarrow R) \end{matrix} \bigg]
\end{split}\label{If}
\end{equation} 
where $\Sigma_{T,LR(RL)}$ is the tunnelling self-energy. The lesser Green's function $G^<=G^r\Sigma^< G^a$~\cite{Samanta1998,Keldysh1964,Gonzales2020,Stefanuccibook2013,Gldecaynote}, where $G^{r/a}$ are the full retarded/advanced Green's function dressed by $\Sigma_{T}$, with all matrices being in the combined Floquet Nambu Keldysh space. Crucially, we obtain the \emph{non-perturbative} Green's functions using matrix inversions instead of an asymptotic perturbative expansion which is divergent~\cite{Bratus1995,Cuevasthesis}. 

For a DC current bias, we impose $I_{k\neq 0}=0$ in Eq. \eqref{If}, and the resulting set of equations determine the voltage across the JJ. The numerical method, inspired by Ref.~\cite{McDonald1976}, is described in Appendix A. 

\begin{figure}[b]
\includegraphics[width=\columnwidth]{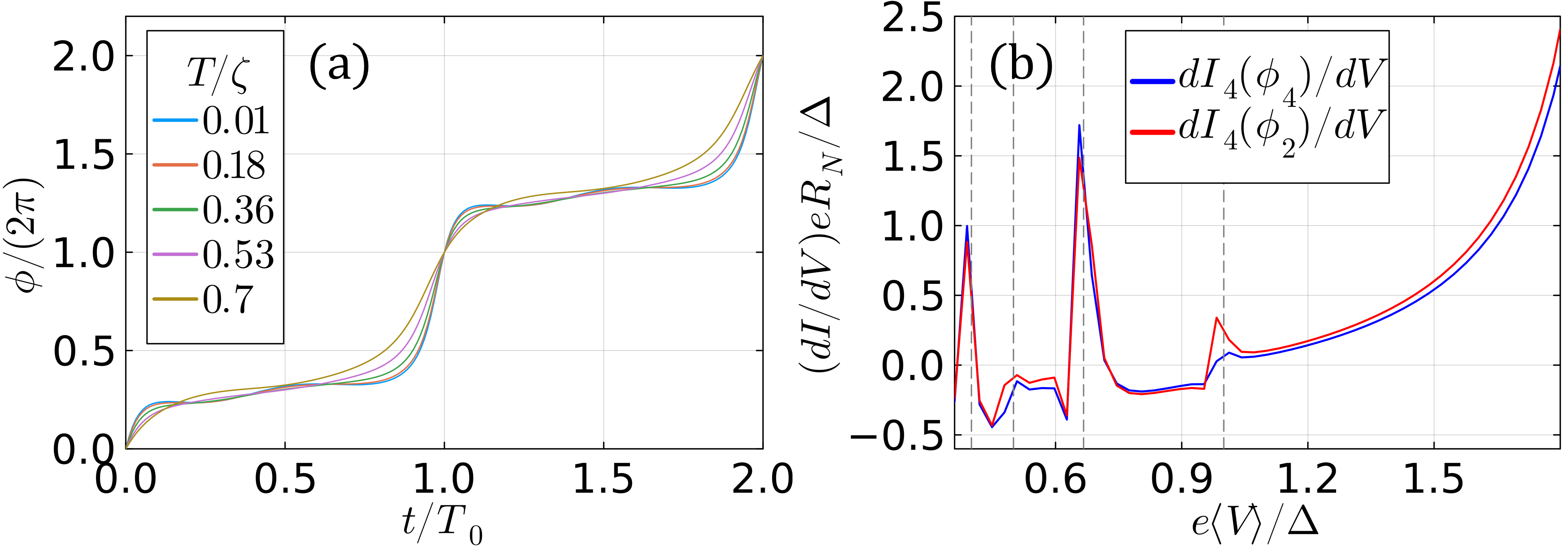}
\caption{(a) Numerically obtained nonperturbative superconducting phase $\phi(t)$ at DC current bias. $\phi(t)$ exhibits steps of height $2\pi$, separated by $T_0$. (b) Differential conductance at $\mathcal{O}(\mathcal{T}^4)$. Even subharmonics appear in both $dI_4(\phi_2)/dV$ and $dI_4(\phi_4)/dV$, where $\phi_{2n}$ is the solution calculated using the $\mathcal{T}^{2n}$ current. Gray lines mark the subharmonics $2\Delta/n$. We use $\Gamma=0.005\Delta$, $\zeta=5\Delta$, and $\mathcal{T}=0.12\zeta$ in (b).}
\label{Fig2} 
\end{figure}

\textit{Numerical results.}---The numerically obtained IVC presented in Fig.~\ref{Fig1}(a,b) show that the even subharmonics are absent for small $\mathcal{T}$, and emerge only with increasing $\mathcal{T}$. This is also evident from the differential conductance presented in Fig.~\ref{Fig1}(c,d), which reveals peaks developing at the even subharmonics with increasing $\mathcal{T}$. At the same time, all subharmonic peaks get commensurately smeared as the gap-edge singularity in the superconducting density of states is renormalised by tunnelling. Additionally, we note that the IVC is hysteretic even in the absence of an external shunting capacitance, arising from the intrinsic quantum capacitance~\cite{Likharevbook}. 

The exact numerical solution for $\phi(t)$ is presented in Fig.~\ref{Fig2}(a). It qualitatively resembles the RCSJ solution for all $\mathcal{T}$, up to small oscillations~\cite{Choi2022}, hinting that the specific functional form of $\phi$ is not crucial for the SGS. Inspired by this observation, we consider two approximated solutions $\phi_2(t)$ and $\phi_4(t)$, obtained using the numerical procedure introduced above, but expanding the current in Eq.~\eqref{If} upto $\mathcal{O}(\mathcal{T}^2)$ and $\mathcal{O}(\mathcal{T}^4)$, respectively. Denoting $I^{(2\nu)}$ as the contribution $\sim (\mathcal{T}/\zeta)^{(2\nu)}$ ~\cite{Yeyati1996,Cuevasthesis,hnote} (see Appendix B), we find that $I_2(\phi_2)=I^{(2)}(\phi_2)$ is simply the WL current and only produces odd subharmonics. However, $I_4(\phi_4)=I^{(2)}(\phi_4)+I^{(4)}(\phi_4)$, along with the corresponding solution $\phi_4$ already generates the missing even subharmonics. Remarkably, we obtain all subharmonics even on plugging $\phi_2$, which is the solution using $I_2$, into $I_4(\phi_2)$ [Fig.~\ref{Fig2}(b)]. Therefore, the SGS do not depend on the specific form of $\phi$, emerging instead from the microscopic tunnel processes captured by the $\mathcal{O}(\mathcal{T}^4)$ current, but overlooked in the $\mathcal{O}(\mathcal{T}^2)$ WL theory.

\textit{Microscopic origin of SGS.}---We explain the microscopic origin of SGS by two complementary pictures. The time-domain picture illustrates that SGS occurs due to self-coupling of retarded current pulses. The frequency-domain picture explains the SGS via energy transfer in microscopic tunnelling processes. Both pictures reveal that the tunnelling of an even number of quasiparticles is essential for the even subharmonics.

We start with the time-domain picture and obtain the voltages exhibiting SGS. Since a DC current bias generates a train of sharp voltage pulses if $\langle V\rangle\le2\Delta/e$, following the discussion in the previous sections and Fig.~\ref{Fig2}(a), we adopt a simple ansatz $\phi(t)$ containing steps of height $2\pi$ separated in time by $T_0=\pi /(e\langle V\rangle)$. Without loss of generality, we calculate the DC component of the pair current $I_s(t)$, as our key results remain unchanged for the quasiparticle currents. Expanding in $\mathcal{T}$, we find $I_s(t) = \sum_{\nu=1}^{\infty}I_s^{(2\nu)}(t)$ where $I_s^{(2\nu)}\sim \mathcal{T}^{2\nu}$ and corresponds to $2\nu e$-charge tunneling. Defining
\begin{eqnarray}
I^{(2\nu)}_s(t) = \Im\{ e^{-2i\nu\phi(t)} \mathcal{I}^{(2\nu)}_s(t) \}, \label{IsTn}
\end{eqnarray}
we first study the $\mathcal{O}(\mathcal{T}^2)$ current~\cite{Choi2022} by analyzing,
\begin{eqnarray}
\mathcal{I}^{(2)}_s(t) && =  \int_{-\infty}^t e^{i[\phi(t)-\phi(t')]/2}\mathcal{K}_s(t-t')dt', \label{Eq:IsNeq(2)(t)}
\end{eqnarray}
where $\mathcal{K}_s(\tau)$ captures the electronic retardation and oscillates at frequency $2\Delta$, stemming from the singular density of states at the superconducting gap-edges~\cite{Harris1976,Choi2022,Lahiri2023} (see Supplemental Material (SM) for details~\cite{SM}). Using integration by parts, we separate $\mathcal{I}^{(2)}_s(t) = I_c^{(2)} + \delta\mathcal{I}^{(2)}_s(t)$ into equilibrium and nonequilibrium quasiparticle-mediated pair tunneling from slow- and fast-varying components of $\phi(t)$, respectively. The latter is given by (see Appendix C)
\begin{equation}
\delta\mathcal{I}^{(2)}_s(t) = ie\int_{-\infty}^t\,V(t')e^{i[\phi(t)-\phi(t')]/2}\mathcal{J}^{(2)}_1(t-t') dt', \label{Eq:deltaIneq(2)(t)}
\end{equation}
where $\mathcal{J}^{(2)}_1(\tau) = \int_{\tau}^{\infty}d\tau'\mathcal{K}_s(\tau)$ and $I_c^{(2)}=\pi\Delta/(2eR_N)$ is the critical current~\cite{SM}. For a small voltage $eV(t)\ll\Delta$, $\delta\mathcal{I}^{(2)}_s$ is negligible, recovering the DC Josephson current $I_s^{(2)}(t) = I_c^{(2)}\sin(\phi(t))$. Focusing on $\phi(t)$ exhibiting a $2\pi$-phase jump at $t=t_0$, corresponding to a sharp voltage pulse at the same time, we find
\begin{equation}
\mathcal{I}^{(2)}_s(t) \approx I_c^{(2)} - 2e^{i[\phi(t)-\phi(t^+_0)]/2}\mathcal{J}^{(2)}_1(t-t_0), \label{Eq:Ineq(2)(t)}
\end{equation}
where $t^+_0\equiv t_0+0^+$ (see Appendix C). The nonequilibrium quasiparticle excitation generated by the voltage pulse relaxes following the energy-time uncertainty, with the corresponding retarded current given by $\mathcal{J}_1^{(2)}$. The fractional phase shift $e^{i[\phi(t)-\phi(t^+_0)]/2}$ in Eq.~\eqref{Eq:Ineq(2)(t)} stems from the dynamical phase of the nonequilibrium tunnelling state with an excess charge of $1e$ relative to the BCS ground state.

To obtain the IVC arising from the DC component of the supercurrent $I_s^{(2)}(t)$, we formulate a superposition principle (SP) wherein we sum the contributions to the current generated by each voltage pulse. By decomposing $V(t)=\sum_{p=0}^{\infty}V_p(t)$ into well-separated voltage pulses $V_p(t)$ at $t_p=-pT_0$, we find the SP
\begin{equation}
\delta\mathcal{I}^{(2)}_s[V] = \sum_{p=0}^{\infty} e^{ie\int_{t_p^+}^t dt'V(t')}\delta\mathcal{I}^{(2)}_s[V_p], \label{Eq:SuperpositionPrinciple}
\end{equation}
which is a quasilinear functional of $V(t)$ with fractional phase shifts~\cite{SM}. 

Applying the SP in Eq.~\eqref{Eq:SuperpositionPrinciple} to a train of voltage pulses corresponding to the ansatz chosen earlier for $\phi(t)$ comprising steps of height $2\pi$ at $t=0,-T_0,\cdots$, we find
\begin{equation}
I_s^{(2)}(t) \approx I_c^{(2)}\sin[\phi(t)] - 2\sum_{p=0}^{\infty}(-1)^p\mathcal{J}^{(2)}_1(t+pT_0). \label{Eq:Is(2)(t)}
\end{equation}
The DC component of the first term matches the IVC of the RCSJ model~\cite{Stewart1968}. The second term leads to SGS at specific voltages when the retarded responses $(-1)^p\mathcal{J}^{(2)}_1(t+pT_0)$ constructively interfere (self-coupling). Using the asymptotic form $\mathcal{J}^{(2)}_1(t)\sim e\Delta\mathcal{T}^2\sin(2t\Delta)\Theta(t)/(t\Delta\zeta^2)$ valid for $t>1/\Delta$, we find the resonance condition of the self-coupling as $2T_0\Delta +\pi = 2n\pi$, yielding the odd subharmonics $e\langle V\rangle = 2\Delta/(2n-1)$. This matches our numerical results in Fig.~\ref{Fig1}(a,c). We highlight that the sign flips $(-1)^p$ in Eq.~\eqref{Eq:Is(2)(t)} stem from excitations with charge $1e$.

Now, we show that the even subharmonics $2\Delta/(2ne)$ arise due to the self-coupling of retarded currents from two quasiparticle excitations with charge $2e$, appearing at order $\mathcal{T}^4$. Considering the pair current as before, we obtain three distinct contributions $\mathcal{I}_s^{(4)}(t)=\sum_{j=1}^3\mathcal{I}_{s,j}^{(4)}(t)$. Without loss of generality, we focus only on $I_{s,1}^{(4)}(t)$ as all the contributions share the same essential features and lead to the same conclusion. As before, we separate $\mathcal{I}^{(4)}_{s,j}(t)=I^{(4)}_{c,j}+\delta\mathcal{I}^{(4)}_{s,j}(t)$~\cite{SM}, and assuming the same step ansatz $\phi(t)$ as in Eq.~\eqref{Eq:Is(2)(t)}, we obtain
\begin{equation}
\delta\mathcal{I}^{(4)}_{s,1}(t) \approx \sum_{k=1}^3\sum_{p=0}^{\infty}e^{ik[\phi(t)-\phi(-pT_0^+)]/2}\mathcal{J}_k^{(4)}(t+pT_0),
\end{equation}
where $\mathcal{J}_{1,2,3}^{(4)}(t)$ correspond to the nonequilibrium quasiparticle-mediated pair tunneling with excess charge $1e$, $2e$, $3e$, respectively. They oscillate at frequency $2\Delta$, similar to $\mathcal{J}_{1}^{(2)}(t)$ in Eq.~\eqref{Eq:Is(2)(t)}. Using the asymptotic expression $\mathcal{J}^{(4)}_2(t)\sim e\Delta\mathcal{T}^4\cos(2t\Delta-3\pi/4)\Theta(t)/(\sqrt{t\Delta}\zeta^4)$ valid for $t>1/\Delta$, we obtain the resonant self-coupling condition $2T_0\Delta = 2n\pi$, resulting in even subharmonics at $e\langle V\rangle= 2\Delta/(2n)$. To conclude, $2e$-excitation-mediated pair tunneling at $\mathcal{O}(\mathcal{T}^4)$ generate the missing even subharmonics. Comprising the tunneling at $\mathcal{O}(\mathcal{T}^2)$- and $\mathcal{O}(\mathcal{T}^4)$ constitutes similar SGS to Fig.~\ref{Fig2}(b). 

Generalising this analysis, we find that supercurrent at $\mathcal{O}(\mathcal{T}^{2\nu})$ with $2\nu e$-charge occurs exclusively via the term $\sim \sin[\nu\phi(t)]$, while quasiparticle-mediated contributions involve $1e,\ldots, (2\nu-1)e$ excitations~\cite{SM}. Hence, the $\mathcal{O}(\mathcal{T}^{2})$ current is a special case, allowing only a single quasiparticle-mediated contribution. It involves odd $1e$ excitation, resulting in the absence of even subharmonics.

Next, we present the frequency-domain picture to illustrate the tunnelling processes. The electronic tunnel pathways defining the current are independent of the biasing scheme. Hence, the conditions for subharmonics, expressed in terms of $W_m$, are universal irrespective of biasing scheme and specific to the order in $\mathcal{T}/\zeta$ at which the current is evaluated. Note that $W_m$ is the dynamical phase acquired by electrons tunelling across the biased junction. Consequently, $W(\omega)$ is the amplitude for the exciting tunnelling quasiparticles by energy $\omega$~\cite{Averin2020,Lahiri2023}. Subharmonics arise only when this excitation satisfies the universal singularity conditions (USCs) specified below. 

The perturbative contribution to the DC current at order $\mathcal{T}^{2n}$ contains $2n$ instances of $\Sigma_T(\omega_j)$ for $j\in 1\ldots 2n$ (expressions at order $\mathcal{T}^2$ and $\mathcal{T}^4$ provided in Appendix B). Each instance of $\Sigma_{T}(\omega_j)$, depending on $W(\omega_j)$ (Appendix A), represents a distinct tunnel event supplying energy $\omega_j$. We require $\sum_j\omega_j=0$ for the DC current. It follows that at any order in $\mathcal{T}$, in general, there are two types of terms: (i) all tunnel events exchange the same magnitude of energy, including usual Andreev processes, and (ii) processes where different tunnel events exchange different energies. For a DC voltage bias $V_{\text{DC}}$, since $W(t)=e^{-ieV_{\text{DC}}t}$ is monochromatic with $W(\omega)=2\pi\delta(\omega-eV_{\text{DC}})$, only type (i) processes exist at all orders in $\mathcal{T}$, with the exchanged energy equalling $eV_{\text{DC}}$. However, for a DC current bias, an AC voltage with mean value $\langle V\rangle$ develops, with $W(\omega)=\sum_{k}W_{(2k-1)}\delta(\omega-(2k-1)e\langle V\rangle)$ containing all \emph{odd} multiples of $e\langle V \rangle$ [explained below Eq. \eqref{If}]. Hence, both processes (i) and (ii) contribute, with the energy exchanged at each tunnel event assuming all these values. Note that, type (ii) is not allowed in the $\mathcal{T}^2$ DC current as there are only two energies satisfying $\omega_1+\omega_2=0$. 

We start with the $\mathcal{T}^2$ DC current (see Appendix D)
\begin{widetext}
\begin{subequations}
\begin{align}
\frac{I_{\text{DC}}^{(2),\text{V bias}}}{4\pi^2 e\mathcal{T}^2}=&{\smallint \limits_{-\infty}^\infty}  \frac{d\omega}{2\pi}A_N\Big(\omega-\mfrac{eV_{\text{DC}}}{2}\Big)A_N\Big(\omega+\mfrac{eV_{\text{DC}}}{2}\Big)\Big[n_F\Big(\omega+\mfrac{eV_{\text{DC}}}{2}\Big)-n_F\Big(\omega-\mfrac{eV_{\text{DC}}}{2}\Big)\Big],\label{Ipertdc2a}\\
\frac{I_{\text{DC}}^{(2),\text{I bias, N}}}{4\pi^2 e\mathcal{T}^2}=&{\sum\displaylimits_{k}\smallint \limits_{-\infty}^\infty} \frac{d\omega}{2\pi} \big(|W_{(2k-1)}|^2-|W_{-(2k-1)}|^2\big)A_N\Big(\omega-\mfrac{(2k-1)e\langle V\rangle}{2}\Big)A_N\Big(\omega+\mfrac{(2k-1)e\langle V\rangle}{2}\Big)\nonumber\\
&\Big[n_F\Big(\omega+\mfrac{(2k-1)e\langle V\rangle}{2}\Big)-n_F\Big(\omega-\mfrac{(2k-1)e\langle V\rangle}{2}\Big)\Big]\label{Ipertdc2b}
\end{align} 
\end{subequations}
\end{widetext}
where $n_F$ is the Fermi distribution, and $A_{N(S)}(\omega)=-(g_{11(12)}^r(\omega)-g_{11(12)}^a(\omega))/(2i\pi)$ is the normal(anomalous) spectral function with the subscripts denoting Nambu indices. For a DC voltage bias, we see from Eq.~\eqref{Ipertdc2a} that resonant transport is only possible at $eV_{\text{DC}}=2\Delta$ by exciting a single quasiparticle from the lower band-edge of one lead to the upper band-edge of the other lead. Thus, it exhibits only the Riedel peak at $eV_{\text{DC}}=2\Delta$, and no subharmonics~\cite{Riedel1964,Werthamer1966}. This defines the type (i) USC. For a DC current bias at order $\mathcal{T}^2$, we have only type (i) processes with both $W$s exchanging the same magnitude of energy $(2k-1)e\langle V\rangle$ $\forall k\in \mathbb{Z}$, which follows from the spectral decomposition of $W(\omega)$. This yields Eq.~\eqref{Ipertdc2b} for the normal current (pair current in Appendix F), from which it is clear that we have the same USC, but we obtain a series of contributions by replacing $eV_{\text{DC}}$ with $(2k-1)e\langle V\rangle$, corresponding to the spectrum of supplied energies. Therefore, we obtain \emph{only} the odd subharmonics at $(2k-1)e\langle V\rangle=2\Delta$.

\begin{figure}[t]
\includegraphics[width=\columnwidth]{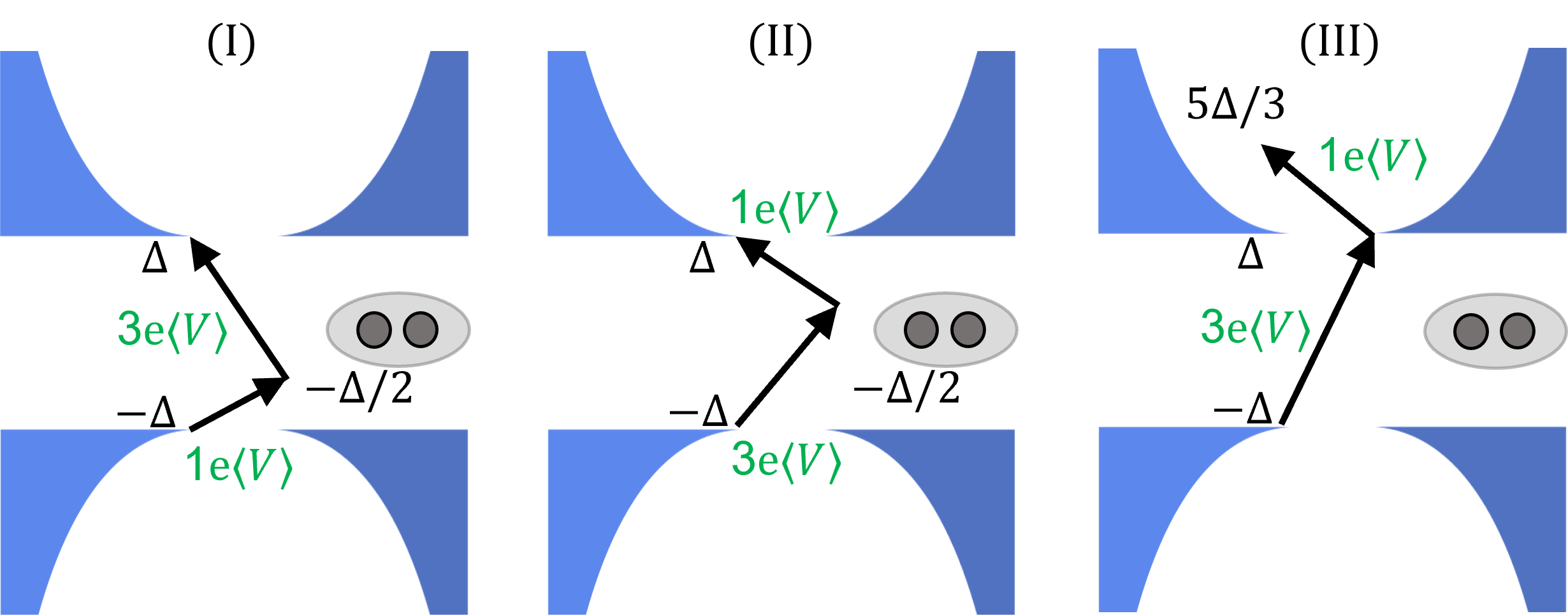}
\caption{(I-III) Illustration of resonant generalised Andreev processes, corresponding to Eq.~\eqref{Ipertdc4b} for $q=1$ and $k=2$, all of them transferring a pair. The density of states is shown in blue, and arrows depict tunnelling quasiparticles. See Tab.~\ref{tab1} for the corresponding resonance conditions. }
\label{Fig3} 
\end{figure}

This argument extends to the $\mathcal{T}^4$ DC current (see Appendix D)
\begin{widetext}
\begin{subequations}
\begin{align}
&\frac{I_{\text{DC}}^{(4),\text{V bias}}}{8\pi^2 e\mathcal{T}^4}={\smallint \limits_{-\infty}^\infty} { \frac{d\omega}{2\pi}A_N(\omega-eV_{\text{DC}})A_N(\omega+eV_{\text{DC}})B_S(\omega)B_S(\omega)\big[n_F(\omega+eV_{\text{DC}})-n_F(\omega-eV_{\text{DC}}) \big]}\label{Ipertdc4a}\\
&\frac{I_{\text{DC}}^{(4),\text{I bias, Andreev}}}{4\pi^2 e\mathcal{T}^4}={\sum\displaylimits_{q,k}\big(|W_{(2q-1)}|^2|W_{(2k-1)}|^2-|W_{-(2q-1)}|^2|W_{-(2k-1)}|^2\big)\smallint \limits_{-\infty}^\infty}  \frac{d\omega}{2\pi} \nonumber\\
&A_N(\omega-(2q-1)e\langle V\rangle)A_N(\omega+(2k-1)e\langle V\rangle)B_S(\omega)B_S(\omega) \big[n_F(\omega+(2k-1)e\langle V\rangle)-n_F(\omega-(2q-1)e\langle V\rangle) \big]+(k\leftrightarrow q) \label{Ipertdc4b}
\end{align} 
\end{subequations}
\end{widetext}
where $B_S=(g_{12}^r(\omega)+g_{12}^a(\omega))/2$. For a DC voltage bias, we see from Eq.~\eqref{Ipertdc4a} showing the first-order Andreev reflection current, that an additional subharmonic occurs at $eV_{\text{DC}}=2\Delta/2$, corresponding to a two-step excitation with each step exciting by energy $eV_{\text{DC}}$~\cite{Bratus1995,Lee2023,MARnote}. This defines the USC for type (i) processes at $\mathcal{T}^4$. For a DC current bias, the DC current mediated by Andreev processes is now given by Eq.~\eqref{Ipertdc4b}. We start with type (i) processes where all excitations have the same magnitude, and thus we restrict to $k=q$ in Eq.~\eqref{Ipertdc4b}. As in the $\mathcal{T}^2$ case, this yields a series of contributions by replacing $eV_{\text{DC}}$ in Eq.~\eqref{Ipertdc4a} with $(2k-1)e\langle V\rangle$, resulting in subharmonics at $(2k-1)e\langle V\rangle=2\Delta/2$. We thus obtain all even subharmonics except for $e\langle V\rangle=2\Delta/(4n)$. Pursuing this strategy to higher orders (Appendix E), we find that for a DC voltage bias, the subharmonic at $eV_{\text{DC}}=2\Delta/n$ appears first at order $ \mathcal{T}^{2n}$~\cite{Bratus1995}.\begin{table}[b]
\renewcommand{\arraystretch}{1.1} % Adjust row height
\setlength{\tabcolsep}{2pt} % Adjust column padding
\begin{tabular}{c c c}
\toprule
\multirow{2}{*}{\textbf{Diagram}} &\hspace{1mm}  \textbf{DC V bias} &\hspace{1mm}  \textbf{DC I bias} \\
\textbf{in Fig.~\ref{Fig3}}  &\hspace{1mm}  \stackanchor{$\Omega = eV_{\mathrm{DC}}$ }{$a=b = 1$ }  &\hspace{1mm}  \stackanchor{$\Omega = e\langle V \rangle$ }{$a, b = \text{odd}$ } \\
\midrule\midrule
(I,II):\ $(|a| + |b|)\Omega = 2\Delta$ &\hspace{1mm} $eV_{\mathrm{DC}} = 2\Delta/2$ &\hspace{1mm} $e\langle V \rangle = 2\Delta/(2n)$ \\ 
(III):\ $a\Omega = 2\Delta$ &\hspace{1mm} $eV_{\mathrm{DC}} = 2\Delta$ &\hspace{1mm} $e\langle V \rangle = 2\Delta/(2n-1)$ \\ 
\bottomrule
\end{tabular}
\caption{USC for both DC voltage and DC current biases, for the generalised Andreev processes in Fig. \ref{Fig3}, corresponding to the diagrams labeled (I-III).}
\label{tab1}
\end{table}  Hence, considering only type (i) process, the subharmonic at $e\langle V\rangle=2\Delta/4n$ appears first at order $\mathcal{T}^{(8n)}$ even for a DC current bias. Remarkably, considering also type (ii) processes, we find that \emph{all} subharmonics, including the ones at $e\langle V\rangle=2\Delta/(4n)$, are obtained already at order $\mathcal{T}^4$. This is evident from Eq.~\eqref{Ipertdc4b}, with the representative physical pathways illustrated in Fig. \ref{Fig3}. Combining two energy exchanges $(2q-1)e\langle V\rangle$ and $(2k-1)e\langle V\rangle$, we obtain a generalised Andreev process which is resonant when the tunnel pathways exploit at the gap edges either the enhanced quasiparticle density of states $A_N$ (Fig.~\ref{Fig3}(I,II)), or the enhanced pairing tendency $B_S$ (Fig.~\ref{Fig3}(III)). This corresponds to $((2q-1)+(2k-1))e\langle V\rangle=2\Delta$ when both normal quasiparticles are extracted from the singular band-edges to form a pair inside the gap, or $(2q-1)e\langle V\rangle=2\Delta$ when the pair forms at the gap edge, thereby defining the type (ii) USC at order $\mathcal{T}^4$. The former is sufficient to generate all subharmonics from a suitable combination of $q$ and $k$, including $e\langle V \rangle=2\Delta/(4n)$ which were previously missed by the type (i) processes. For instance, $e\langle V \rangle=2\Delta/2$ is obtained from $q=1$ and $k=1$, while $e\langle V \rangle=2\Delta/4$ is obtained from $q=1$ and $k=2$. A similar analysis applies to the pure normal and pure pair currents (Appendix F). 

Hence, unlike a DC voltage bias, the wide spectrum of energies supplied by the AC voltage corresponding to a DC current bias exhausts \emph{all} subharmonics already at order $\mathcal{T}^4$. Our treatment provides a unified approach for both, as summarised in Table \ref{tab1}. Finally, we note that the higher subharmonics are progressively suppressed as they rely on: (a) higher harmonics in $W(\omega)$, which are increasingly attenuated, and (b) higher order currents $I_{\text{DC}}^{(2n)}$, which are suppressed by the small factor $(\mathcal{T}/\zeta)^{(2n)}$.
\\

\textit{Conclusion}.---We provide a microscopic theoretical solution for DC current-biased Josephson junctions with arbitrary junction transparencies. Transcending the WL theory, which is the leading second-order term in tunnel coupling and fails to match the experimental subharmonic gap structure (SGS), we find that the fourth-order term generates correct SGS for high transparencies. We identify the two-quasiparticle processes responsible for capturing the previously missing subharmonics. 

\begin{acknowledgements} A. L. and B. T. acknowledge support by the W\"urzburg-Dresden Cluster of Excellence ct.qmat, EXC2147, project-id 390858490, and the DFG (SFB 1170). S.-J. C. acknowledges support by the research grant of Kongju National University in 2024. A.L. acknowledges helpful discussions with Jacob F. Steiner.
\end{acknowledgements}

\textit{Author contributions}.---A.L. developed the numerical Floquet calculations, and the frequency-domain perturbative explanation based on the generalised Andreev processes. S.-J. C. formulated the time-domain perturbative explanation. All authors contributed equally to conceptualising the work, and writing the manuscript.

\let\oldaddcontentsline\addcontentsline% Store \addcontentsline
\renewcommand{\addcontentsline}[3]{}% Make \addcontentsline a no-op

\section{End Matter}
\textit{Appendix A}.---Here, we describe the numerical Floquet technique. The Floquet Keldysh Green's functions are expanded as~\cite{Gavenskythesis,Kuhn2023},
\begin{equation}
G(t,t')=\sum_{m,n} \int_0^\Omega \frac{d\omega}{2\pi} e^{-i(\omega+m\Omega)t}e^{i(\omega+n\Omega)t'}G_{mn}(\omega),
\end{equation}
where $\Omega=2\pi/T=e\langle V\rangle$ is the Floquet frequency , and $G_{mn}(\omega+k\Omega)=G_{(m+k)(n+k)}(\omega)$. The Dyson equation becomes $G_{mn}(\omega)=g_{mn}(\omega)+g_{mk}(\omega)\Sigma_{kk'}(\omega)G_{k'n}(\omega)$, where the Green's functions are in Floquet Nambu space, and $g_{mn}(\omega)=g(\omega+m\Omega)\delta_{mn}$ is the bare Green's function. 

The self-energy contains three terms: (i) $\Sigma_{\mathcal{T}}$ arising from $H_{\mathcal{T}}$, given by $\Sigma^{r/a}_{T,LR,,m-n}={\Sigma^{r/a}}^*_{T,RL,n-m}=\int_0^T H_\mathcal{T}(\tau)e^{i(m-n)\Omega \tau} d\tau/T=-\mathcal{T}[W_{m-n}(\tau_0+ \tau_3)/2 + W_{-(m-n)}^*(\tau_0- \tau_3)/2]$, where $W_{a}$ is the Floquet transform of $W(t)=e^{-i\phi(t)/2}$, and $\tau$ denotes the Pauli matrices in Nambu space. The lesser component vanishes~\cite{Cuevas1996}. (ii) $\Sigma^{r/a}_\Gamma=\pm i\Gamma/2$ and $\Sigma^{<}_\Gamma=-i\Gamma f(\omega)$, where $f(\omega)$ is the Fermi function. Apart from aiding numerical convergence, it accounts for the broadening/lifetime arising from, e.g., relaxation to the continuum, electron-phonon interactions, etc~\cite{Lahiri2023}. (iii) $\Sigma_\zeta$, resulting from coupling the central device region composed of single-site leads ~\cite{ssitenote} to infinite superconducting reservoirs~\cite{Samanta1998}. Specifically, $\Sigma^{r/a/<}_\zeta=(-\zeta/2)^2\tau_3g^{r/a/<}\tau_3$ where $g^{r/a/<}$ is the boundary Green's function~\cite{Peng2017,Zazunov2016,Samanta1998,Cuevas1996,grnote}. 

Following Ref.~\cite{McDonald1976}, we consider $4N_F+1$ Floquet modes $W_n$ for the factor $W(t)=e^{-i\phi(t)/2}$, ranging from $-2N_F\Omega$ through $2N_F\Omega$. $W(t)$ satisfies $W(t+T_0)=-W(t)$, and thus only the \emph{odd} harmonics $W_{(2n-1)}\neq 0$. This corresponds to $4N_F$ variables since $W$ is complex. The Floquet modes of the Green's functions and the current $I$ lie in the range $-N_F\Omega$ to $N_F\Omega$. The first $2N_F$ equations to be solved are obtained by imposing $I_{2k}=0$. $I_{2k-1}$ vanish naturally for conventional superconductors. We obtain $2N_F-1$ additional equations from the condition $W(t)W(t)^*=1$. The final equation is $\phi(t=0)=0$. These are solved for different values of $\Omega=e\langle V\rangle$. This amounts to obtaining the AC voltage with the mean value $\langle V\rangle$ which generates a DC current. Since the quasiparticles repeatedly tunnel (Andreev process) until they escape into the continuum, the required $N_F$ scales as $2\Delta/eV$. In practice, we use $N_F=34$ in Fig. \ref{Fig1}.

The exact current reduces to the WL current~\cite{Werthamer1966} for small $\mathcal{T}$. We expand $G^<=g^<+\delta{G^<}^{(1)}$ with $g^<=g^r\Sigma^<g^a\sim\mathcal{O}(\mathcal{T}^0)$, and $\delta {G^<}^{(1)}= (g^r\Sigma_Tg^<+g^<\Sigma_Tg^a)$ being the leading $\mathcal{O}(\mathcal{T})$ correction. Note that $g^{r/a/<}$ include the broadening $\Gamma$ with $g^<=g^r\Sigma^<_\Gamma g^a$, but not tunnelling (see \cite{Gldecaynote}). The corresponding $\mathcal{O}(\mathcal{T}^2)$ current $I(t)=e\Re\ \mathbf{Tr}\int_{-\infty}^\infty dt' \tau_3 H_{T}(t) \big( g^r(t-t') H_T(t')g^<(t'-t)+g^<(t-t') H_T(t')g^a(t'-t) \big)$ is the WL current~\cite{Werthamer1966,Lahiri2023}.
\\

\textit{Appendix B}.---Using the Langreth rules~\cite{Haugbook}, the $\sim\mathcal{T}^\nu$ contribution to the lesser Green's function is given by,
\begin{equation}
\delta {G^<}^{(\nu)}= \sum_{i=1}^\nu \bigg(\prod_{j=1}^{i-1} g_j^r\Sigma^r_j\bigg) g_i^<\bigg(\prod_{k=i+1}^\nu\Sigma^a_{k}g_{k}^a\bigg).\label{delG}
\end{equation}
Hence, the current $I=e\mathbf{tr}[\Sigma^T_{LR}\delta G^<_{RL}]-(L\leftrightarrow R)$, at order $\mathcal{T}^{2}$ and $\mathcal{T}^{4}$, are given by
\begin{subequations}
\begin{align}
&\frac{I_{\text{DC}}^{(2)}}{e\mathcal{T}^2}=\begingroup\textstyle\int\endgroup\limits_{-\infty}^\infty \hspace{-0.4em} \frac{d\omega}{2\pi} \prod\limits_{j=1}^2 \frac{d\omega_j}{2\pi} \delta\big(\sum_{l=1}^2\omega_j\big) \mathbf{tr}\big[ \tau_3 \Sigma^T_{LR}(\omega_1)\nonumber \\
& \big(g(\omega+\omega_2)\Sigma^T_{RL}(\omega_2)g(\omega)\big)^< -(L\leftrightarrow R) \big], \label{Ipertdc24a}\\
&\frac{I_{\text{DC}}^{(4)}}{e\mathcal{T}^4}=\begingroup\textstyle\int\endgroup\limits_{-\infty}^\infty \hspace{-0.4em} \frac{d\omega}{2\pi}\prod\limits_{j=1}^4 \frac{d\omega_j}{2\pi} \delta\big(\sum_{l=1}^4\omega_j\big)\mathbf{tr}\big[ \tau_3 \Sigma^T_{LR}(\omega_1) \nonumber\\
&\big(g(\omega+\omega_2+\omega_3+\omega_4)\Sigma^T_{RL}(\omega_2)g(\omega+\omega_3+\omega_4)\nonumber\\
&\Sigma^T_{LR}(\omega_3)g(\omega+\omega_4)\Sigma^T_{RL}(\omega_4)g(\omega)\big)^< - (L\leftrightarrow R) \big].\label{Ipertdc24b}
\end{align} 
\end{subequations}\\

\textit{Appendix C}.---First, we derive $\mathcal{I}^{(2)}_s(t)=I^{(2)}_c + \delta\mathcal{I}^{(2)}_s(t)$ stated in the main text. Applying integration by parts to Eq.~\eqref{Eq:IsNeq(2)(t)}, we obtain
\begin{eqnarray}
\mathcal{I}^{(2)}_s(t) && = e^{\frac{i}{2}\phi(t)}\left\{\left[e^{-\frac{i}{2}\phi(t')}\mathcal{J}^{(2)}_1(t-t')\right]\Big|_{t'=-\infty}^{t'=t} \right. \nonumber\\
&& \left. + ie\int_{-\infty}^t V(t') e^{-\frac{i}{2}\phi(t')} \mathcal{J}^{(2)}_1(t-t')dt'\right\},
\end{eqnarray}
where $\mathcal{J}^{(2)}_1(t-t')$ is a primitive function of $\mathcal{K}_s(t-t')$ (see Ref.~\cite{SM} for the definition). To decompose $\mathcal{I}^{(2)}_s(t)$ into two contributions from slow- and fast-varying $\phi(t)$, we consider a small voltage $eV(t)\ll\Delta$, and use the boundary condition $\mathcal{J}^{(2)}_1(\infty)=0$ and $\mathcal{J}^{(2)}_1(0)=I_c^{(2)}$, which hold if $\mathcal{J}^{(2)}_1(t-t') \equiv \int_{t-t'}^{\infty}\mathcal{K}_s(\tau)d\tau$. Consequently, we have $\mathcal{I}^{(2)}_s(t) = I_c^{(2)} + \delta\mathcal{I}^{(2)}_s(t)$. Indeed, $\mathcal{J}^{(2)}_1(0) =\pi\Delta/(2eR_N)$, yielding the critical current at $\mathcal{O}(\mathcal{T}^2)$. 

Now, we obtain Eq.~\eqref{Eq:Ineq(2)(t)}. We consider an abrupt phase jump at $t=t_0$ over a slowly-varying phase $\phi_0(t)$
\begin{equation}
\phi(t) = \left\{
\begin{array}{ll}
\phi_0(t), & \quad t<t_0, \\
\phi_0(t) + \frac{\delta\phi}{\delta t}(t-t_0), & \quad t_0 < t < t_0+\delta t, \\
\phi_0(t) + \delta\phi, & \quad t_0+\delta t<t,
\end{array}\right.
\end{equation}
where $\phi_0'(t) \ll 2e\Delta$. Then, the nonzero value of Eq.~\eqref{Eq:deltaIneq(2)(t)} is obtained only for $t>t_0+\delta t$ as follows.
\begin{align}
&\delta\mathcal{I}^{(2)}_s(t)\nonumber \\
&= \frac{i e^{\frac{i\phi(t)}{2}}\delta\phi}{2\delta t}\int_{t_0}^{t_0+\delta t}e^{-\frac{i\delta\phi(t'-t_0)}{2\delta t}}e^{-\frac{i\phi_0(t')}{2}}\mathcal{J}^{(2)}_1(t-t') dt' \nonumber \\
&= -e^{\frac{i\phi(t)}{2}}(e^{-\frac{i\phi(t_0+\delta t)}{2}} - e^{-\frac{i\phi(t_0)}{2}})\mathcal{J}^{(2)}_1(t-t_0) + \mathcal{O}(\delta t).
\end{align}
where we obtain the last equation by expanding $e^{-i\phi_0(t')/2}\mathcal{J}^{(2)}_1(t-t')$ at $t_0$ in powers of $\delta t$. In the main text, we present the lowest order of $\delta t$, and the Heaviside function $\Theta(t-t_0)$ is absorbed into the asymptotic expression of $\mathcal{J}^{(2)}_1(t-t_0)$.\\

\textit{Appendix D}.---
Here we sketch the derivation of the analytical expressions for the current given by Eqs.~\eqref{Ipertdc2a}, \eqref{Ipertdc2b}, \eqref{Ipertdc4a}, and \eqref{Ipertdc4b}, with the last two representing the generalised Andreev process. Recall that for a DC voltage bias, $W(t)=\exp(-ieV_{\text{DC}}t)$, whereas for a DC current bias we use $W(t)=\sum_m W_{(2m-1)}\exp(-i(2m-1) e\langle V\rangle t)$. We start with the current given by Eqs.~\eqref{Ipertdc24a} and \eqref{Ipertdc24b}, and perform the Nambu traces. For the generalised Andreev process at order $\mathcal{T}^4$ in Eq.~\eqref{Ipertdc24b}, we retain only  terms with two normal and two anomalous Green's funtions, and ensure that either two $W$s or $W^*$s appear successively. We also restrict to the case that the right(left) leads generate only anomalous(normal) Green's functions, which holds when the left lead has a higher electrochemical potential. To match the stated result, we define the normal(anomalous) spectral functions $A_{N(S)}=-(g^r_{11(12)}-g^a_{11(12)})/(2\pi i)$, and $B_S=(g^r_{12}+g^a_{12})$, and finally consider $\Gamma\ll \Delta$ which yields the simplifying identity $A_{\alpha}(\omega)B_{\beta}(\omega)\to 0$ where $\alpha,\beta=N/S$. In this limit, $A(B)$ is non-zero only out(in)-side the superconducting gap.\\

\begin{figure}[!htb]
\includegraphics[width=\columnwidth]{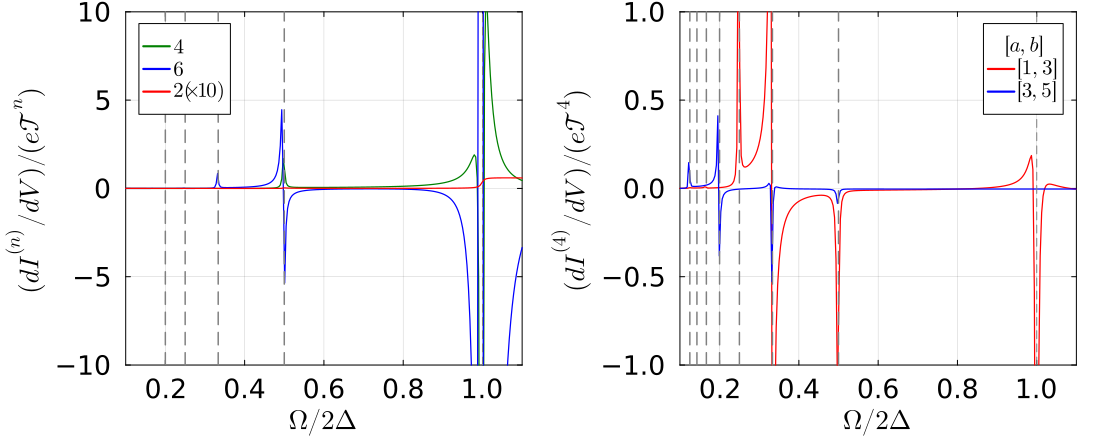}
\caption{(Left) $(dI_{\text{DC}}^{(n)}/dV)/(e\mathcal{T}^n)$ for type-(i) processes, where all tunnel events exchange the same energy $\Omega$. The singularity at $\Omega=2\Delta/n$ appears first in $I_{\text{DC}}^{(2n)}$. (Right) Type-(ii) processes in $(dI_{\text{DC}}^{(4)}/dV)/(e\mathcal{T}^4)$ with two different energies exchanged $\Omega_1=a\Omega$ and $\Omega_2=b\Omega$ (see legend), with $\Omega=e\langle V\rangle$ for a DC current bias. The dashed vertical lines denote all subharmonics. The USC follow: $\Omega=2\Delta/a$, $2\Delta/b$, and $2\Delta/(|a|+|b|)$ as mentioned in Tab. \ref{tab1}. These arise from the generalised Andreev process and the pure pair current (Appendix F). There is an additional subharmonic at $2\Delta/||a|-|b||$, which comes from the pure normal current (Appendix F).}
\label{Fig4}
\end{figure}

\textit{Appendix E}.---We show the universal singularity conditions in currents at various orders of $\mathcal{T}$ in Fig.~\ref{Fig4}.\\ 

\begin{figure}[!htb]
\includegraphics[width=\columnwidth]{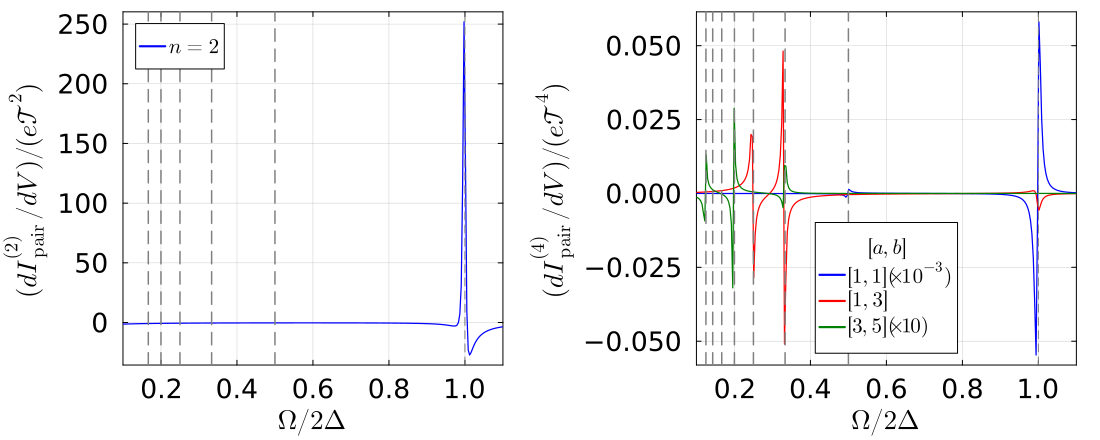}
\caption{Same as Fig.~\ref{Fig4}, but for pure pair currents. (left) $(dI_{\text{DC}}^{(2)}/dV)/(e\mathcal{T}^2)$ for type (i) process. (right) $(dI_{\text{DC}}^{(4)}/dV)/(e\mathcal{T}^4)$ for type (ii) processes with two different energies exchanged $\Omega_1=a\Omega$ and $\Omega_2=b\Omega$ (see legend), with $\Omega=e\langle V\rangle$ for a DC current bias. We find the same USC as stated in the main text, and mentioned in Fig. \ref{Fig4} except for the one at $\Omega=2\Delta/(|a|-|b|)$.}
\label{Fig5}
\end{figure}
\begin{figure}[!htb]
\includegraphics[width=0.7\columnwidth]{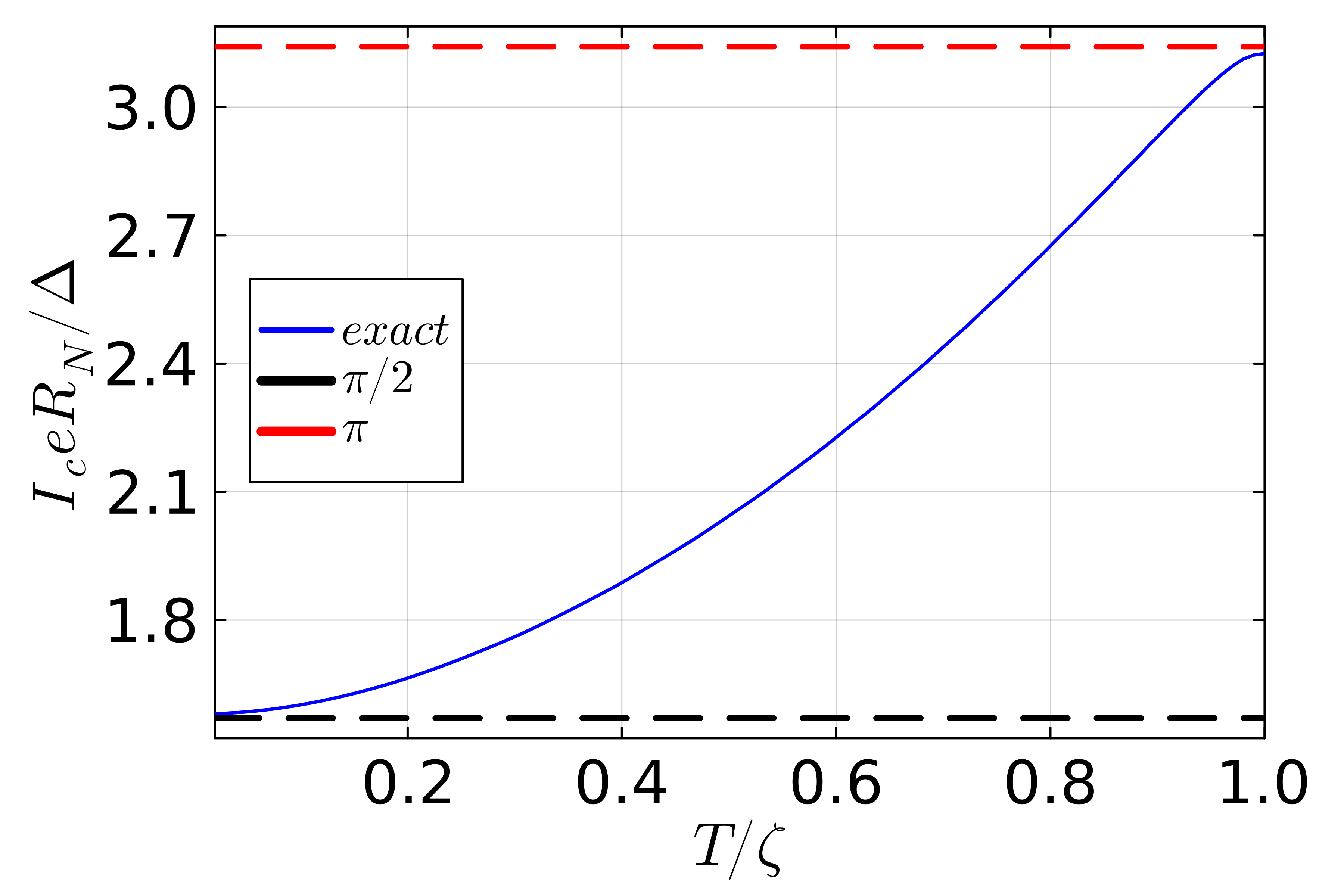}
\caption{We plot the ``figure of merit" $I_c eR_n/\Delta$, where $I_c$ is the critical current, $R_N$ is the normal state resistance. The two limits: $\pi/2$ in the tunnel limit, and $\pi$ in the transparent limit, are shown with dashed lines. }
\label{Fig6}
\end{figure}

\textit{Appendix F}.---While we have shown the generalised Andreev processes in Fig. \ref{Fig3} (see also Ref.~\cite{MARnote}), there is also a pure normal and a pure pair current at $\mathcal{T}^4$. They are obtained from Eqs.~\eqref{Ipertdc24a} and~\eqref{Ipertdc24b} by retaining terms containing only the normal and anomalous Green's functions, respectively. 

(a) Pure pair current: For a DC voltage bias $eV_{\text{DC}}$, for which $W(\omega)=2\pi\delta(\omega-eV_{\text{DC}})$, it can be checked from Eqs.~\eqref{Ipertdc24a} and~\eqref{Ipertdc24b} that the DC pure pair term at any order $\mathcal{T}^{2\nu}$ contains a string of either $2\nu$ $W$s or $2\nu$ $W^*$s. It thus vanishes as the net energy supplied equals $\pm 2\nu eV_{\text{DC}}$ (creating an AC current with frequency $2\nu eV_{\text{DC}}$), while the DC current requires net zero energy. However, for a DC current bias, which effectively corresponds to an AC voltage as described in the main text, there is a non-zero DC pure pair current at all orders, even for finite voltages~\cite{McDonald1976,Choi2022}. This arises as now $W$ contains \emph{both positive and negative} odd multiples of $e\langle V\rangle$. Hence, even though we have a string of only $W$s or $W^*$s, the net energy can add to zero. We provide the analytical result at order $\mathcal{T}^2$ in Eq.~\eqref{Ipertdcapair2}. It depends on the quasiparticle density and the pairing tendency ($B_S=(g^r_{12}+g^a_{12})/2$), both of which are singular near the superconducting gap edges, resulting in odd subharmonics at $(2k-1)e\langle V\rangle=2\Delta$. The $\mathcal{T}^4$ pair current in Fig.~\ref{Fig5} shows similar subharmonics as the generalised Andreev process (Fig.~\ref{Fig3} and Tab.~\ref{tab1}).

(b) Pure normal current: At all orders, it depends only on the normal density of states $A_N$, similar to the $\mathcal{T}^2$ result given by Eq.~\eqref{Ipertdc2b}. At $\mathcal{T}^4$, while it can show the subharmonics at $e\langle V\rangle=2\Delta/a$, $2\Delta/b$ ($a,b$ are odd) as the quasiparticle can directly jump across the gap already in the first one of the two jumps (similar to Fig.~\ref{Fig3}(III)), it cannot show the subharmonic at $e\langle V\rangle=2\Delta/(|a|+|b|)$. This is because, starting at $-\Delta$, after the first jump the quasiparticle necessarily ends up in the sub-gap region $-\Delta+\text{max}(|a|,|b|)e\langle V\rangle$ (similar to Fig.~\ref{Fig3}(I,II)). However, it has a new subharmonic at $e\langle V\rangle=2\Delta/||a|-|b||$ ($a,b$ are odd), where the quasiparticle at $-\Delta$ first absorbs $\text{max}(|a|,|b|)e\langle V\rangle$ to reach $-\Delta+\text{max}(|a|,|b|)e\langle V\rangle>\Delta$, and then loses $|b|e\langle V\rangle$ to drop down to $+\Delta$. This bypasses the sub-gap region, and is thus allowed. We find this in Fig.~\ref{Fig4}, which shows the full current, and hence it includes the pure normal component. Note that this $e\langle V\rangle=2\Delta/||a|-|b||$ subharmonic is not found in the generalised Andreev process and the pure pair current as the intermediate step involves $B_S$ instead of $A_{N/S}$, with $B_{N/S}(|\omega|>\Delta)\to 0$.\\

\textit{Appendix G}.--- We show the quantity $I_c eR_n/\Delta$ in Fig. \ref{Fig6}, which spans the range $\pi/2$ in the tunnel limit, through $\pi$ for unit transparency (corresponding to $\mathcal{T}=\zeta$)~\cite{Golubov2004}.

\begin{widetext}
\begin{align}
\frac{I_{\text{DC, pair}}^{(2),\text{I bias}}}{4\pi^2 e\mathcal{T}^2}=&{\sum\displaylimits_{k}\int\limits_{-\infty}^\infty}  \frac{d\omega}{2\pi}\mfrac{\big(W^*_{-(2k-1)}W^*_{(2k-1)}-W_{(2k-1)}W_{-(2k-1)}\big)}{\pi}\Big[ A_S\Big(\omega-\mfrac{(2k-1)e\langle V\rangle}{2}\Big)B_S\Big(\omega+\mfrac{(2k-1)e\langle V\rangle}{2}\Big)\nonumber\\
&n_F\Big(\omega-\mfrac{(2k-1)e\langle V\rangle}{2}\Big) + A_S\Big(\omega+\mfrac{(2k-1)e\langle V\rangle}{2}\Big)B_S\Big(\omega-\mfrac{(2k-1)e\langle V\rangle}{2}\Big)n_F\Big(\omega+\mfrac{(2k-1)e\langle V\rangle}{2}\Big)\ \Big].\label{Ipertdcapair2}
\end{align} 
\end{widetext}


\begin{thebibliography}{70}
\bibitem{ssitenote} Since we assume uniform superconducting order parameter and phase in the leads, we may transfer all but the single site neighbouring the barrier region into the semi-infinite reservoirs.
\bibitem{grnote} The surface Green's function is $g^r=(-(w+i0)\tau_0+\Delta\tau_1)/(\zeta\sqrt{\Delta^2-(w+i0)^2})$, where $\tau_\nu$ denote the Pauli matrices in the Nambu space. It satisfies $(g^r)^{-1}=(g^r_0)^{-1}-\Sigma^r_\zeta$, where $g_0^r=[(w+i0)\tau_0-(-\mu\tau_3+\Delta\tau_1)]^{-1}$.
\end{thebibliography}
\end{document}

\title{Supplemental Material for ``Microscopic Theory of DC Current-Biased Josephson Junctions with Arbitrary Transparencies''}
\author{Aritra Lahiri}
\email[]{aritra.lahiri@uni-wuerzburg.de}
\affiliation{Institute for Theoretical Physics and Astrophysics,
University of W\"urzburg, D-97074 W\"urzburg, Germany}
\author{Sang-Jun Choi}
\email[]{aletheia@kongju.ac.kr}
\affiliation{Department of Physics Education, Kongju National University, Gongju 32588, Republic of Korea}
\author{Björn Trauzettel}
\affiliation{Institute for Theoretical Physics and Astrophysics, University of W\"urzburg, D-97074 W\"urzburg, Germany}
\affiliation{W\"urzburg-Dresden Cluster of Excellence ct.qmat, Germany}
\date{\today}

\begin{abstract}
In this Supplemental Material, we present the standard mean-field Hamiltonian of superconductors used in the time-domain analysis, various anomalous bare Green's functions, the derivation of the superposition principle of nonequilibrium quasiparticle tunneling, the detailed calculations of $\delta\mathcal{I}_s^{(4)}(t)$, and the discussion of $\delta\mathcal{I}_s^{(2\nu)}(t)$. Here, we reintroduce $\hbar$ when it is necessary for clarity.
\end{abstract}
\maketitle

\section{Standard mean-field Hamiltonian of superconductors}

We provide the standard mean-field Hamiltonian of $s$-wave superconductors and tunneling Hamiltonian adopted for the time-domain analysis in the main text. The Hamiltonian of the superconductors on left and right $H_L$ and $H_R$ are presented in the momentum space $k$,
\begin{eqnarray}
H_L &=& \sum_{k,\sigma = \uparrow, \downarrow} \left[ (E_k-\mu_L)c_{k,\sigma,L}^\dagger c_{k,\sigma,L} + \Delta_L c_{k,\sigma,L}^\dagger c_{-k,-\sigma,L}^\dagger + \Delta_L^* c_{-k,-\sigma,L} c_{k,\sigma,L} \right], \\
H_R &=& \sum_{k,\sigma = \uparrow, \downarrow} \left[ (E_k-\mu_R)c_{k,\sigma,R}^\dagger c_{k,\sigma,R} + \Delta_R c_{k,\sigma,R}^\dagger c_{-k,-\sigma,R}^\dagger + \Delta_R^* c_{-k,-\sigma,R} c_{k,\sigma,R} \right],
\end{eqnarray}
where $\sigma$ is the spin index of electrons, and $-\sigma$ denotes the opposite direction of the spin. $\mu_L$ and $\mu_R$ are the chemical potential of left and right superconductors, respectively. In the main text, we focus on a symmetric junction with the same superconducting gap $\Delta_L = \Delta_R = \Delta$. The tunneling Hamiltonian is also presented in the momentum space,
\begin{eqnarray}
H_{\mathcal{T}} = \sum_{k,q,\sigma}\left[ \mathcal{T}e^{-i\phi(t)/2}c_{k,\sigma,R}^\dagger c_{q,\sigma,L} + \mathcal{T}^*e^{i\phi(t)/2}c_{q,\sigma,L}^\dagger c_{k,\sigma,R} \right].
\end{eqnarray}
The tunneling matrix element $\mathcal{T}$ is related to the transition probability of an electron with momentum $q$ in the left lead to tunnel to the momentum state $k$ in the right lead. It is valid to use W.K.B approximation and introduce the momentum-independent tunneling matrix element $\mathcal{T}$, for voltages comparable to the superconducting gap or the junctions with a quantum point contact~\cite{Barone}. Since only powers of $|\mathcal{T}|^2$ appear in the calculation of the current, we assume $\mathcal{T}$ is real without loss of generality.

\section{Various anomalous bare Green's functions}
Here we provide the anomalous bare Green's functions. For convenience, we absorb a factor of $\mathcal{T}$ into the bare Green's function, as they appear later while calculating the perturbative expansion of the dressed Green's functions.
\begin{eqnarray}
f^r(t) &=& \sum_k f^r_{k,q}(t) =  -\frac{\Delta}{2\hbar}\frac{\mathcal{T}}{\zeta}\Theta(t)J_0\left(\frac{\Delta|t|}{\hbar}\right), \\
f^a(t) &=& \sum_k f^a_{k,q}(t) =  -\frac{\Delta}{2\hbar}\frac{\mathcal{T}}{\zeta}\Theta(-t)J_0\left(\frac{\Delta|t|}{\hbar}\right), \\
f^>(t) &=& \sum_k f^>_{k,q}(t) =  -\frac{\Delta}{4\hbar}\frac{\mathcal{T}}{\zeta}\sgn(t)J_0\left(\frac{\Delta|t|}{\hbar}\right) + i\frac{\Delta}{4\hbar}\frac{\mathcal{T}}{\zeta} Y_0\left(\frac{\Delta|t|}{\hbar}\right), \\
f^<(t) &=& \sum_k f^<_{k,q}(t) =  \frac{\Delta}{4\hbar}\frac{\mathcal{T}}{\zeta} J_0\left(\frac{\Delta|t|}{\hbar}\right) + i\frac{\Delta}{4\hbar}\frac{\mathcal{T}}{\zeta} Y_0\left(\frac{\Delta|t|}{\hbar}\right),
\end{eqnarray}
where we evaluate the real-time Green's functions at zero temperature using the BCS approximation of the pair density of states in the wide-band limit~\cite{Barone},
\begin{eqnarray}
p(\omega) = \frac{\Delta}{\sqrt{\omega^2-\Delta^2}}\Theta(|\omega|-\Delta).
\end{eqnarray}
Here, $J_0$ and $Y_0$ are Bessel functions of the first and second kind, respectively. The normal resistance is inversely proportional to $\mathcal{T}^2$ as $1/R_N=4\pi e^2 N_F^2\mathcal{T}^2/\hbar$ with the density of states at Fermi energy $N_F=1/(2\pi\zeta)$. We consider a symmetric junction with the superconducting gap $\Delta$ and bandwidth $\zeta$. Moreover, the normal conductance is $G_n=1/R_N=2e^2\mathbb{T}/h$ with the transmission probability across the junction $\mathbb{T} = 4(\mathcal{T}/\zeta)^2/(1+\mathcal{T}^2/\zeta^2)^2$. We note that the well-known formula $1/R_N=4\pi e^2 N_F^2\mathcal{T}^2/\hbar^3$ has a typographical error.

\section{Superposition Principle of nonequilibrium quasiparticle tunneling}

We show the superposition principle of nonequilibrium quasiparticle tunneling in detail. Notice that, for any given time $t$, $\delta \mathcal{I}^{(2)}_s$ can be seen as a functional of a time-dependent voltage $V(t)$,
\begin{eqnarray}
\delta \mathcal{I}^{(2)}_s[V]  &\equiv&  ie\int_{-\infty}^t\,dt'V(t')e^{i[\phi(t)-\phi(t')]/2}\mathcal{J}^{(2)}_1(t-t') \nonumber\\
 &=& ie\int_{-\infty}^t\,dt'V(t')e^{ie\int_{t'}^t d\tau V(\tau)}\mathcal{J}^{(2)}_1(t-t').
\end{eqnarray}
We highlight the following features: (i) $\delta \mathcal{I}^{(2)}_s[V]$ depends solely on the voltage bias $V(t)$ which is a physical observable and the kernel $\mathcal{J}^{(2)}_1(t-t')$ which includes the microscopic details of the junction, (ii) $\delta \mathcal{I}^{(2)}_s[V]$ may disappear at weak voltages, and (iii) the time-integration depends only on the time duration over which the voltage $V(t)$ is significant.  

First, let us consider that the voltage $V(t)$ is decomposed into a sum of disjoint voltages $V(t)=V_1(t)+V_2(t)$ having mutually exclusive supports, i.e., either $V_1(t)=0$ or $V_2(t)=0$ for any time $t$. Without loss of generality, we assume that $V_1(t)$ precedes $V_2(t)$ in time. Then, there exists a time $t_1$ lying between their supports such that $V_1(t_1)=0$ and $V_2(t_1)=0$. Now, we show the quasilinearity of the functional $\delta \mathcal{I}^{(2)}_s[V]$
\begin{eqnarray}
&& \delta \mathcal{I}^{(2)}_s[V_1+ V_2] = ie\int_{-\infty}^t\,dt'V_1(t')e^{ie\int_{t'}^t d\tau V(\tau)}\mathcal{J}^{(2)}_1(t-t') - ie\int_{\infty}^t\,dt'V_2(t')e^{ie\int_{t'}^t d\tau V(\tau)}\mathcal{J}^{(2)}_1(t-t') \nonumber\\
&&\qquad = ie\int_{-\infty}^{t_1}\,dt'V_1(t')e^{ie\int_{t'}^t d\tau V_1(\tau)}e^{ie\int_{t'}^t d\tau V_2(\tau)}\mathcal{J}^{(2)}_1(t-t')  - ie\int_{t_1}^t\,dt'V_2(t')e^{ie\int_{t'}^t d\tau V_1(\tau)}e^{ie\int_{t'}^t d\tau V_2(\tau)}\mathcal{J}^{(2)}_1(t-t') \nonumber\\
&&\qquad = ie\int_{-\infty}^{t_1}\,dt'V_1(t')e^{ie\int_{t'}^t d\tau V_1(\tau)}e^{\bcancel{ie\int_{t'}^{t_1} d\tau V_2(\tau)}+ie\int_{t_1}^t d\tau V_2(\tau)}\mathcal{J}^{(2)}_1(t-t') \nonumber\\
&& \qquad\quad - ie\int_{t_1}^t\,dt'V_2(t')e^{\bcancel{ie\int_{t'}^t d\tau V_1(\tau)}}e^{ie\int_{t'}^t d\tau V_2(\tau)}\mathcal{J}^{(2)}_1(t-t') \nonumber\\
&&\qquad =  ie\,e^{ie\int_{t_1}^t d\tau V_2(\tau)}\int_{-\infty}^{t}\,dt'V_1(t')e^{ie\int_{t'}^t d\tau V_1(\tau)}\mathcal{J}^{(2)}_1(t-t') - ie\int_{-\infty}^t\,dt'V_2(t')e^{ie\int_{t'}^t d\tau V_2(\tau)}\mathcal{J}^{(2)}_1(t-t') \nonumber\\
&&\qquad =  e^{ie\int_{t_1}^{t}d\tau V(\tau)} \delta \mathcal{I}^{(2)}_s[V_1] + \delta \mathcal{I}^{(2)}_s[V_2].  \label{SMEq:linearity}
\end{eqnarray}

Now, we generalise this anaysis by considering a voltage $V(t)$ comprising a series of disjoint pulses $V_p(t)$, $V(t)=\sum_{p=0}^N V_p(t)$. As before, there exist particular moments $t_p$ lying between the supports of $V_{p}(t)$ and $V_{p+1}(t)$ such that $V_{p}(t_p)=0$ and $V_{p+1}(t_p)=0$. Equivalently, $V_p(t)$ is nonzero for $t_{p-1}<t<t_p$. For convenience, we consider $t_{-1}$ at which $V_0(t<t_{-1})=0$.
\begin{eqnarray}
&& \delta \mathcal{I}^{(2)}_s\left[\sum_{p=0}^N V_p(t)\right] \nonumber\\
&&\qquad = ie\sum_{p=0}^N\int_{-\infty}^t\,dt'V_p(t')e^{ie\int_{t'}^t d\tau V(\tau)}\mathcal{J}^{(2)}_1(t-t') = ie\sum_{p=0}^N\int_{t_{p-1}}^{t_p}\,dt'V_p(t')e^{ie\int_{t'}^t d\tau V(\tau)}\mathcal{J}^{(2)}_1(t-t') \\
&&\qquad = ie\sum_{p=0}^N\int_{t_{p-1}}^{t_p}\,dt'V_p(t')e^{ie\int_{t'}^{t_p}d\tau V(\tau)+ ie\int_{t_{p}}^{t}d\tau V(\tau)}\mathcal{J}^{(2)}_1(t-t') \\
&&\qquad = ie\sum_{p=0}^N\int_{t_{p-1}}^{t_p}\,dt'V_p(t')e^{ie\int_{t'}^{t_p}d\tau V_p(\tau)+ ie\int_{t_{p}}^{t}d\tau V(\tau)}\mathcal{J}^{(2)}_1(t-t') \\
&&\qquad = ie\sum_{p=0}^N e^{ie\int_{t_{p}}^{t}d\tau V(\tau)} \int_{t_{p-1}}^{t_p}\,dt'V_p(t')e^{ie\int_{t'}^{t}d\tau V_p(\tau)}\mathcal{J}^{(2)}_1(t-t')\\
&&\qquad = ie\sum_{p=0}^N e^{ie\int_{t_{p}}^{t}d\tau V(\tau)} \int_{-\infty}^{t}\,dt'V_p(t')e^{ie\int_{t'}^{t}d\tau V_p(\tau)}\mathcal{J}^{(2)}_1(t-t') = \sum_{p=0}^N e^{ie\int_{t_{p}}^{t}d\tau V(\tau)} \delta \mathcal{I}^{(2)}_s[V_p].
\end{eqnarray}
Sending $N\rightarrow\infty$ and $t_p\rightarrow T_p^+$, we obtain the superposition principle provided in the main text.

We note that the quasilinearity holds approximately when the overlap between the voltage pulses is negligibly small. In other words, for any given time $t$, either $V_1(t) \ll V_2(t)$ or $V_2(t) \ll V_1(t)$. This quasilinearity allows us to determine the supercurrent at time $t$ in terms of a superposition of the retarded responses of voltage pulses in the past. Furthermore, it has the significant implication of determining the excessive charge in nonequilibrium tunneling. The superposition principle of supercurrent in nonequilibrium tunneling is a rather surprising result, in light of the nonlinear nature of equilibrium Josephson effect, $I_c\sin[\phi_1(t)+\phi_2(t)]\neq I_c\sin[\phi_1(t)]+I_c\sin[\phi_2(t)]$.

\section{Calculations of $\mathcal{I}^{(4)}_s(t)$}
Here, we calculate the supercurrent $I_s^{(4)}(t)$ at order $\mathcal{T}^4$, dividing it into equilibrium and nonequilibrium components, with the latter mediated by intermediate nonequilibrium quasiparticle tunneling. Applying the Langreth rule to obtain lesser and greater Green's functions at order $\mathcal{T}^4$, the total supercurrent is obtained as $I_s^{(4)}(t) = I_{s,1}^{(4)}(t) + I_{s,2}^{(4)}(t) + I_{s,3}^{(4)}(t)$ and $I_{s,j}^{(4)}(t) = \Im\{e^{-2i\phi(t)}\mathcal{I}^{(4)}_{s,j}(t)\}$ with
%\begin{eqnarray}
%\mathcal{I}_{s,1}(t) &=& e^{3i\phi(t)/2}\Pint{t}{t'} f^>(t-t')\Pint{t'}{t'_1}f^r(t'-t_1')\Pint{t_1'}{t_2'}f^r(t_1'-t_2')f^<(t_2'-t) \nonumber \\
%&& - e^{3i\phi(t)/2}\Pint{t}{t'} f^<(t-t')\Pint{t'}{t'_1}f^r(t'-t_1')\Pint{t_1'}{t_2'}f^r(t_1'-t_2')f^>(t_2'-t).
%\end{eqnarray}
\begin{eqnarray}
\mathcal{I}^{(4)}_{s,1}(t) &=& -4ei\,e^{3i\phi(t)/2}\Pint{t}{t'}\Pint{t'}{t_1} \Pint{t_1}{t_2} \nonumber\\
&& [f^>(t-t')f^r(t'-t_1)f^r(t_1-t_2)f^<(t_2-t) - f^<(t-t')f^r(t'-t_1)f^r(t_1-t_2)f^>(t_2-t) ], \\
\mathcal{I}^{(4)}_{s,2}(t) &=& -4ei\,e^{3i\phi(t)/2}\Pint{t}{t'}\Pint{t'}{t_1} \Pint{t}{t_2} \nonumber\\
&& [f^>(t-t')f^r(t'-t_1)f^<(t_1-t_2)f^a(t_2-t) - f^<(t-t')f^r(t'-t_1)f^>(t_1-t_2)f^a(t_2-t) ], \\
\mathcal{I}^{(4)}_{s,3}(t) &=& -4ei\,e^{3i\phi(t)/2}\Pint{t}{t'}\ \Pint{t}{t_2'} \Pint{t_2'}{t'_1} \nonumber\\
&& [f^>(t-t')f^<(t'-t_1)f^a(t_1-t_2)f^a(t_2-t) - f^<(t-t')f^>(t'-t_1)f^a(t_1-t_2)f^a(t_2-t) ],
\end{eqnarray}
where the integrand is a purely imaginary function. For example, the integrand of $\mathcal{I}_{s,1}(t)$ is
\begin{eqnarray}
&&f^>(t-t')f^r(t'-t_1')f^r(t_1'-t_2')f^<(t_2'-t) - f^<(t-t')f^r(t'-t_1')f^r(t_1'-t_2')f^>(t_2'-t) \\
&& = -\frac{i \pi^2 \Delta^4}{8e^4\hbar^2 R_N^2} J_0\left(\frac{(t-t') \Delta }{\hbar }\right) J_0\left(\frac{(t'-t_1) \Delta }{\hbar }\right) J_0\left(\frac{(t_1-t_2) \Delta }{\hbar }\right)  Y_0\left(\frac{(t-t_2) \Delta }{\hbar }\right) \nonumber \\ 
&& \quad -\frac{i \pi^2 \Delta^4}{8e^4\hbar^2 R_N^2} Y_0\left(\frac{(t-t') \Delta }{\hbar }\right) J_0\left(\frac{(t'-t_1) \Delta }{\hbar }\right) J_0\left(\frac{(t_1-t_2) \Delta }{\hbar }\right)  J_0\left(\frac{(t-t_2) \Delta }{\hbar }\right). \nonumber
\end{eqnarray}
We use the normal resistance $R_N$,
\begin{eqnarray}
\frac{1}{R_N} = \frac{e^2}{\pi\hbar}\frac{\mathcal{T}^2}{\zeta^2},
\end{eqnarray}
to facilitate a convenient comparison with the benchmark set by the product $I_c R_N$.

Before proceeding with the calculation of $\mathcal{I}_{s,j}(t)$, we introduce shorthand notations of repeatedly appearing integrals,
\begin{eqnarray}
I[f(t_1,\sla{{\color{gray}t_2}}{{\color{magenta}1}},t_3,\cdots)]_{_{\color{magenta}1}}({\color{blue}t}) &\equiv& {\color{magenta}\int_{-\infty}^{{\color{blue}t}}} d{\color{gray}t_2} f(t_1,{\color{gray}t_2},t_3,\cdots), \\
P_n[f(t_1,\sla{{\color{gray}t_2}}{{\color{magenta}1}},t_3,\cdots)]_{_{\color{magenta}1}}({\color{blue}t}) &\equiv& {\color{magenta}\int_{-\infty}^{{\color{blue}t}}} d{\color{gray}t_2} e^{-in\phi({\color{gray}{t_2}})/2} f(t_1,{\color{gray}t_2},t_3,\cdots), \\
V_n[f(t_1,\sla{{\color{gray}t_2}}{{\color{magenta}1}},t_3,\cdots)]_{_{\color{magenta}1}}({\color{blue}t}) &\equiv& i\frac{ne}{\hbar}{\color{magenta}\int_{-\infty}^{{\color{blue}t}}} d{\color{gray}t_2} V({\color{gray}{t_2}})e^{-in\phi({\color{gray}{t_2}})/2} f(t_1,{\color{gray}t_2},t_3,\cdots),
\end{eqnarray}
where $V(t) = \hbar\phi'(t)/(2e)$. The slashed variables refer to dummy variables of the integral. Since the integrations of $\mathcal{I}^{(4)}_{s,1}(t)$ are nested, slash notations should be distinguished with super- and subscripts to match dummy variables to the associated integrals. Once a variable is slashed by integrating it out, the remaining integral is a function of the unslashed variables. We provide the central identities involving the shorthand notations,
\begin{eqnarray}
P_n[e^{-i\sla{\phi({\color{gray}t_2'})}{{\color{magenta}1}}/2}f(t_1,\sla{{\color{gray}t_2'}}{{\color{magenta}1}},t_3,\cdots)]_{_{\color{magenta}1}}({\color{blue}t}) &=& P_{n+1}[f(t_1,\sla{{\color{gray}t_2'}}{{\color{magenta}1}},t_3,\cdots)]_{_{\color{magenta}1}}({\color{blue}t}), \\
P_n[f(t_1,\sla{{\color{gray}t_2'}}{{\color{magenta}1}},t_3,\cdots)]_{_{\color{magenta}1}}({\color{blue}t}) &=& e^{-in\phi({\color{blue}t})/2} I[f(t_1,\sla{{\color{gray}t_2'}}{{\color{magenta}1}},t_3,\cdots)]_{_{\color{magenta}1}}({\color{blue}t}) + V_n[ I[f(t_1,\sla{{\color{gray}t_2'}}{{\color{magenta}1}},t_3,\cdots)]_{_{\color{magenta}1}}\sla{({\color{gray}t_2})}{{\color{magenta}2}} ]_{_{\color{magenta}2}}({\color{blue}t}),
\end{eqnarray}
where we obtain the last relation using integration by parts. 

Finally, applying the above identities, we derive
\begin{eqnarray}
&& \mathcal{I}^{(4)}_{s,1}(t) \nonumber\\
&& = P_1\left[ f^>(t-\sla{t'}{1}) \left( e^{-i\phi\sla{(t')}{1}}A(\sla{t'}{1},t) + B(\sla{t'}{1},t) + e^{-i\phi\sla{(t')}{1}/2}C(\sla{t'}{1},t) + D(\sla{t'}{1},t)\right)\right]_{_1}(t) - \left(f^> \leftrightarrow f^<\right), \nonumber\\
&& = \Iint{t}{t'} f^>(t-t')A(t',t) \nonumber\\
&& + e^{i\phi(t)/2} \left\{\Iint{t}{t'}f^>(t-t')C(t',t) \right\} \nonumber\\
&& + e^{i\phi(t)}\left\{ \Iint{t}{t'}f^>(t-t')B(t',t) + \Iint{t}{t'}f^>(t-t')D(t',t)\right\} \nonumber\\
&& + e^{3i\phi(t)/2} \left\{ \VVVint{t}{t'}\Iint{t'}{t''}f^>(t-t'')A(t'',t) \right. \nonumber\\
&& + \VVint{t}{t'}\Iint{t'}{t''}f^>(t-t'')C(t'',t) \nonumber\\
&& + \left. \Vint{t}{t'}\Iint{t'}{t''}f^>(t-t'')B(t'',t) + \Vint{t}{t'}\Iint{t'}{t''}f^>(t-t'')D(t'',t) \right\}, \nonumber\\
&&  - \left(f^> \leftrightarrow f^<\right) \label{SMEq:Inje4s1(t)}
\end{eqnarray}
where the terms in the curly brackets vanish for small voltages $V(t)\ll 2\Delta/e$,
\begin{eqnarray}
A(t',t) &\equiv& -4ei\Iint{t'}{t_1'} f^r(t'-t_1')\Iint{t_1'}{t_2'}f^r(t_1'-t_2')f^<(t_2'-t), \\
B(t',t) &\equiv&  -4ei\VVint{t'}{t_1}\Iint{t_1}{t_1'}f^r(t'-t_1')\Iint{t_1'}{t_2'}f^r(t_1'-t_2')f^<(t_2'-t), \\
C(t',t) &\equiv&  -4ei\Iint{t'}{t_1'}f^r(t'-t_1')\Vint{t_1'}{t_2}\Iint{t_2}{t_2'}f^r(t_1'-t_2')f^<(t_2'-t), \\
D(t',t) &\equiv&  -4ei\Vint{t'}{t_1}\Vint{t_1}{t_1'}f^r(t'-t_1')\Iint{t_1'}{t_2}\Iint{t_2}{t_2'}f^r(t_1'-t_2')f^<(t_2'-t).\nonumber\\
\end{eqnarray}
Noticing that $A(t',t)$ is free from the time-dependent voltage, we have separated the equilibrium supercurrent portion with a critical current $I_{c,1}^{(4)}$ at $\mathcal{T}^4$-order,
\begin{eqnarray}
I_{c,1}^{(4)} &=& -4e\Iint{t}{t'} \Iint{t'}{t_1'} \Iint{t_1'}{t_2'} [f^>(t-t')f^r(t'-t_1')f^r(t_1'-t_2')f^<(t_2'-t) - \left(f^> \leftrightarrow f^<\right)] \nonumber\\
&=& -4e \int_0^\infty d\tau \int_0^\infty d\tau_1 \int_0^\infty d\tau_2 [f^>(\tau)f^r(\tau_1)f^r(\tau_2)f^<(-\tau-\tau_1-\tau_2) - \left(f^> \leftrightarrow f^<\right)]. \label{SMEq:Ic(4)}
\end{eqnarray}
Using integration by substitution, it is shown that $I_{c,1}^{(4)}$ is independent of $t$.

Notice that the voltage function appears only once in the multiple integral of $B(t',t)$ and $C(t',t)$ so that the superposition principle of nonequilibrium quasiparticle tunneling can be applied. Especially, if the voltage consists of a train of pulses centered at $T_0,T_1,\cdots$, corresponding to sharp $2\pi$-jumps in the Josephson phase, we can find concise expressions
\begin{eqnarray}
B(t',t) &\approx& \frac{4e\hbar}{\pi\Delta}\sum_{p=0}^{\infty}e^{-i\phi(T_p+0^+)}f^r(t'-T_p)\Iint{T_p}{t_2'}f^r(T_p-t_2')f^<(t_2'-t), \\
C(t',t) &\approx&  8e\sum_{p=0}^{\infty}e^{-i\phi(T_p^+)/2}\Iint{t'}{t_1'}f^r(t'-t_1')\Iint{T_p}{t_2'}f^r(t_1'-t_2')f^<(t_2'-t),
\end{eqnarray}
where we have used the fact that the width of the voltage pulses in time is $\hbar/(\pi\Delta)$, valid for $eV\ll\Delta$. 

We provide the terms containing a single instance of the voltage function, along with the prefactors $e^{i\phi(t)/2}$, $e^{i\phi(t)}$, $e^{3i\phi(t)/2}$, each corresponding to the supercurrent accompanied by the excess quasiparticle charge $1e$, $2e$, $3e$
\begin{eqnarray}
&& \mathcal{I}^{(4)}_{s,1}(t) \nonumber\\
&& = I^{(4)}_{c,1} - 2\sum_{p=0}^{\infty}e^{i[\phi(t)-\phi(T_p^+)]/2}\mathcal{J}_1^{(4)}(t-T_p)  - \sum_{p=0}^{\infty}e^{i[\phi(t)-\phi(T_p^+)]}\mathcal{J}_2^{(4)}(t-T_p) - 2\sum_{p=0}^{\infty}e^{3i[\phi(t)-\phi(T_p^+)]/2}\mathcal{J}_3^{(4)}(t-T_p) \nonumber\\
&& \qquad + \text{(terms containing two and three voltage pulses)}, \nonumber\\
&& = I^{(4)}_{c,1} - 2\sum_{p=0}^{\infty}(-1)^p[\mathcal{J}_1^{(4)}(t-T_p)+\mathcal{J}_3^{(4)}(t-T_p)]  - \sum_{p=0}^{\infty}\mathcal{J}_2^{(4)}(t-T_p) + \text{(terms containing two and three voltage pulses)}, \nonumber\\ \label{SMEq:Ineq(4)(t)}
\end{eqnarray}
where the last equation is derived assuming the voltage pulses corresponding to $2\pi$-phase jumps. We note that the terms accompanied by an odd number of elementary charge excitations is associated with the sign flip $(-1)^p$. We explicitly show that the nonequilibrium retarded responses are merely shifted in time by $T_p$ with the same functional behaviour,
\begin{eqnarray}
&& \mathcal{J}_1^{(4)}(t-T_p) \equiv -4ei\Iint{t}{t'}f^>(t-t') \Iint{t'}{t_1'}f^r(t'-t_1')\Iint{T_p}{t_2'}f^r(t_1'-t_2')f^<(t_2'-t)- \left[f^> \leftrightarrow f^<\right], \nonumber\\
&& \qquad\qquad\,\,\, =  -4ei\int_0^\infty d\tau f^>(\tau) \int_0^\infty d\tau_1 f^r(\tau_1) \int_0^\infty d\tau_2 f^r(t-T_p-\tau-\tau_1+\tau_2)f^<(T_p-t-\tau_2)- \left[f^> \leftrightarrow f^<\right],   \\
&& \mathcal{J}_2^{(4)}(t-T_p) \equiv -4ei\frac{\hbar}{\pi\Delta}\Iint{t}{t'} f^>(t-t') f^r(t'-T_p)\Iint{T_p}{t_2'}f^r(T_p-t_2')f^<(t_2'-t)- \left[f^> \leftrightarrow f^<\right], \nonumber\\
&& \qquad\qquad\,\,\, = -4ei\frac{\hbar}{\pi\Delta}\int_{0}^{\infty}d\tau f^>(\tau) f^r(t-T_p-\tau)\int_{0}^{\infty}d\tau_2 f^r(\tau_2)f^<(T_p-t-\tau_2)- \left[f^> \leftrightarrow f^<\right], \label{SMEq:B(t-Tn)}\\
&& \mathcal{J}_3^{(4)}(t-T_p) \equiv -4ei\Iint{T_p}{t'}f^>(t-t')\Iint{t'}{t_1'} f^r(t'-t_1')\Iint{t_1'}{t_2'}f^r(t_1'-t_2')f^<(t_2'-t)- \left[f^> \leftrightarrow f^<\right]  \nonumber\\
&& \qquad\qquad\,\,\, = -4ei\int_{0}^{\infty}d\tau f^>(t-T_p-\tau) \int_{0}^{\infty}d\tau_1 f^r(\tau_1)\int_{0}^{\infty}d\tau_2 f^r(\tau_2)f^<(T_p-t-\tau-\tau_1-\tau_2)- \left[f^> \leftrightarrow f^<\right].
\end{eqnarray}
We note that $\mathcal{J}_{1,2,3}^{(4)}(t-T_p)$ oscillate with the frequency $2\Delta/\hbar$, similar to the case at $\mathcal{T}^2$-order, and decay in time. The retarded response mediated by $2e$-tunneling $\mathcal{J}_2^{(4)}$, which plays an essential role for even integer fraction of SGS, has an asymptotic expression for large argument $1<\Delta(t-T_p)/\hbar$,
\begin{equation}
\mathcal{J}_2^{(4)}(t-T_p) \approx \frac{\pi^2\hbar\Delta}{e^3R_N^2}\frac{\cos\left[\frac{2\Delta(t-T_p)}{\hbar}-\frac{3\pi}{4}\right]}{\sqrt{\Delta(t-T_p)/\hbar}}\Theta(t-T_p). \label{SMEq:asympB(t-Tn)}
\end{equation}
While the complicated analytical formulae of $\mathcal{I}^{(4)}_{s,2}(t)$ and $\mathcal{I}^{(4)}_{s,3}(t)$ are not provided here, we note that they share the same qualitative properties as described here for $\mathcal{I}^{(4)}_{s,1}(t)$. 

Now, we discuss the properties of terms depending on two instances of voltages in the past, such as terms including $D(t',t)$. We find a concise expression for a train of sharp $2\pi$-phase jumps,
\begin{eqnarray}
D(t',t) &\approx&  4\sum_{n=1}^{\infty}\sum_{m>n}^{\infty}e^{-i\phi(T_n^+)/2}e^{-i\phi(T_m^+)/2}f^r(t'-T_m)\Iint{T_m}{t_2}\Iint{t_2}{t_2'}f^r(T_m-t_2')f^<(t_2'-t).
\end{eqnarray}
Using this, we simplify the term with the prefactor $e^{i\phi(t)}$ in Eq.~\eqref{SMEq:Inje4s1(t)} into
\begin{eqnarray}
&& e^{i\phi(t)}\Iint{t}{t'}f^>(t-t')D(t',t) \nonumber\\
&& = 4\sum_{n=1}^{\infty}\sum_{m>n}^{\infty}e^{i[\phi(t)-\phi(T_n^+)]/2}e^{i[\phi(t)-\phi(T_m^+)]/2}\Iint{t}{t'}f^>(t-t')f^r(t'-T_m)\Iint{T_m}{t_2}\Iint{t_2}{t_2'}f^r(T_m-t_2')f^<(t_2'-t) \nonumber\\
&& = 4\sum_{n=1}^{\infty}\sum_{m>n}^{\infty}e^{i[\phi(t)-\phi(T_n^+)]/2}e^{i[\phi(t)-\phi(T_m^+)]/2} \int_{0}^{\infty}d\tau f^>(\tau)f^r(t-T_m-\tau) \int_{0}^{\infty}d\tau_1 \int_{0}^{\infty}d\tau_2 f^r(\tau_1+\tau_2)f^<(T_m-t-\tau_1-\tau_2) \nonumber\\
&& = 4\sum_{n=1}^{\infty}\sum_{m>n}^{\infty}e^{i[\phi(t)-\phi(T_n^+)]/2}e^{i[\phi(t)-\phi(T_m^+)]/2}\mathcal{D}(t-T_m).
\end{eqnarray}
While the tunneling process from this term is seemingly determined by two voltage pulses, in fact, the retarded response is generated by a single voltage pulse with the other one attaching the Josephson phase. Hence, some combination of voltage pulses will preserve the sign of $\mathcal{D}(t)$, thereby contributing to the even integer fractions in the subharmonic gap structure. A similar analysis may be performed for all terms in Eq.~\eqref{SMEq:Inje4s1(t)}. However, in order to maintain the simple line of logic explaining the subharmonic gap structure at order $\mathcal{T}^2$, let us focus on the integrals in which the voltage appears only once. We leave the complete analysis for our future work.

\section{Discussion of $\mathcal{I}^{(2\nu)}_s(t)$}
The analysis of $\mathcal{I}^{(4)}_s(t)$ at $\mathcal{T}^4$-order is readily applied to that of $\mathcal{T}^{2\nu}$-order ($\nu=1,2,\cdots$). At $\mathcal{T}^{2\nu}$-order,
\begin{eqnarray}
\mathcal{I}^{(2\nu)}_s(t) = -4ei\,e^{i\frac{2\nu-1}{2}\phi(t)} \prod_{n=1}^{2\nu-1}\int_{-\infty}^{\infty} dt_n e^{-i\phi(t_n)/2}[\text{sum of products of $2\nu$ anomalous bare Green's functions}].
\end{eqnarray}
Similar to Eq.~\eqref{SMEq:Ic(4)}, the term independent of a voltage function yields the critical equilibrium supercurrent at $\mathcal{T}^{2\nu}$-order, and the remaining terms give rise to nonequilibrium excitations with the phase factors of $\phi(t)/2, \phi(t), \cdots, (2\nu-1)\phi(t)/2$, which correspond to excess $1e,2e,\cdots,(2\nu-1)e$-charge excitations. Hence, we may argue that the only tunneling process with effective charge $e^*=2\nu e$ at $\mathcal{T}^{2\nu}$-order is the equilibrium supercurrent of $I_s^{(2\nu)}(t) = I_c^{(2\nu)}\sin[\nu\phi(t)]$, while the nonequilibrium tunneling is mediated with excess charges $e,\cdots,(2\nu-1)e$ and $2\nu e$-excitation is not allowed. As a result, the supercurrent at $\mathcal{T}^{2\nu}$-order can be written as
\begin{eqnarray}
I_s^{(2\nu)}(t) = I_c^{(2\nu)}\sin[\nu\phi(t)] + \sum_{q=1}^{2\nu-1}\sum_{p=0}^\infty\Im\left\{ e^{iq[\phi(t)-\phi(T_p^+)/2}\mathcal{J}_{q}^{(2\nu)}(t-T_p) \right\}.
\end{eqnarray}

Indeed, the supercurrent at $\mathcal{T}^2$-order is an exceptional case allowing only the odd number of elementary charge tunneling of $1e$. 
\begin{eqnarray}
I_s^{(2)}(t) = I_c^{(2)}(t)\sin[\phi(t)] + \sum_{p=0}^\infty\Im\left\{ e^{i[\phi(t)-\phi(T_p^+)]/2}\mathcal{J}_{1}^{(4)}(t-T_p) \right\}.
\end{eqnarray}
The nonequilibrium supercurrent $\mathcal{I}^{(2)}_s(t)$ with only $1e$-tunneling is the physical origin of the missing even integer fraction of subharmonic gap structure. Since the $2e$-tunneling at $\mathcal{T}^2$-order allows only Cooper pair tunneling, while quasiparticle tunneling is allowed only via $1e$-tunneling, the even integer fractions are not obtained in the SGS. However, $2e$-tunneling at $\mathcal{T}^4$-order is mediated via nonequilibrium quasiparticle tunneling, thereby generating the even subharmonics at $2\Delta/e(\text{even})$. It can be easily seen that at order $\mathcal{T}^{2\nu}$ for any $\nu>1$, we have both odd and even $ne$-tunneling, and consequently all odd and even subharmonics are obtained in the SGS.